\documentclass{aa}
\usepackage{graphicx}
\usepackage{txfonts}
\usepackage{longtable}
\usepackage{natbib}
\bibpunct{(}{)}{;}{a}{}{,} 
\begin{document}
   \title{XMM-\textit{Newton} observations of the Galactic globular clusters \object{NGC~2808}~and~\object{NGC~4372}}


   \author{M. Servillat
          \inst{1}
          \and
          N. A. Webb
          \inst{1}
          \and
          D. Barret
          \inst{1}
          }


   \institute{CESR, Universit\'e Paul Sabatier, CNRS -- 9 avenue du Colonel Roche, 31400 Toulouse, France \\
              \email{mathieu.servillat@cesr.fr}
             }

   \date{Received 20 July 2007 / Accepted 28 November 2007}

 
  \abstract
   {}
   {Galactic globular clusters harbour binary systems that are detected as faint X-ray sources. These close binaries are thought to play an important role in the stability of the clusters by liberating energy and delaying the inevitable core collapse of globular clusters. The inventory of close binaries and their identification is therefore essential.}
   {We present XMM-\textit{Newton} observations of two Galactic globular clusters: \object{NGC~2808} and \object{NGC~4372}. We use X-ray spectral and variability analysis combined with ultra-violet observations made with the XMM-\textit{Newton} optical monitor and published data from the Hubble Space Telescope to identify sources associated with the clusters. We compare the results of our observations with estimates from population synthesis models.}
   {Five sources out of 96 are likely to be related to \object{NGC~2808}. Nine sources are found in the field of view of \object{NGC~4372}, none being located inside its half-mass radius. We find one quiescent neutron star low mass X-ray binary candidate in the core of \object{NGC~2808}, and propose that the majority of the central sources in \object{NGC~2808} are cataclysmic variables. An estimation leads to $20\pm10$ cataclysmic variables with luminosity above $4.25\times10^{31}\mathrm{~erg~s}^{-1}$. Millisecond pulsars could also be present in the core of \object{NGC~2808}, and some sources outside the half-mass radius could possibly be linked to the cluster.}
   {}


   \titlerunning{XMM-\textit{Newton} observations of \object{NGC~2808}~and~\object{NGC~4372}}
   \authorrunning{Servillat et al.}
   \maketitle
%

\section{Introduction}

XMM-\textit{Newton} and \textit{Chandra} X-ray observatories are currently revealing more and more faint X-ray sources in globular clusters (GCs) thanks to their high sensitivity and high angular resolution respectively \citep[e.g.][]{Webb+04,WWB06,Heinke+03,Heinke+06}. Thirteen bright X-ray sources with ${L_X>10^{36}\mathrm{~erg~s}^{-1}}$ are found in the $152$ known Galactic GCs. These are neutron star low-mass X-ray binaries (LMXBs) showing type I X-ray bursts \mbox{\citep[e.g.][]{1983adsx.conf...41L}}. Numerous faint X-ray sources with ${L_X<10^{34.5}\mathrm{~erg~s}^{-1}}$ have been shown to be a variety of objects, mainly binaries such as LMXBs in quiescence (qLMXBs), cataclysmic variables (CVs), active binaries (ABs), or millisecond pulsars (MSPs), the probable progenity of LMXBs. 
These objects have been classified through multiwavelength analysis: qLMXBs are usually identified by their soft blackbody-like X-ray spectra \mbox{\citep[e.g.][]{GBW03,GBW03b}}, CVs can be confirmed by their blue, variable optical counterpart \mbox{\citep[e.g.][]{Webb+04}}, ABs by their main-sequence, variable optical counterparts \mbox{\citep[e.g.][]{EGHG03}}, and MSPs by their radio counterpart \mbox{\citep[e.g.][]{GHEM01}}.

It is clear that through mass segregation, heavy objects such as binaries are concentrated towards the core of GCs \citep{LG82}. Therefore, we expect the majority of X-ray binaries, which are more massive than the mean stellar mass, to be located inside the half-mass radius.
\citet{HAS07} showed from simulations that the fraction of primordial binaries destroyed in the core by a variety of processes is balanced by the combination of mass-segregation and creation of new binaries in exchange interactions, leading to a marked increase of the binary fraction in the central regions. 
Outside the half-mass radius, the primordial binary fraction is well preserved \citep{HAS07} which could
explain the presence of close binaries located outside of the half-mass radius, as a CV in \object{M~22} \citep{Pietrukowicz+05}. \citet{DAmico+02} and \citet{CMP03} also discussed the case of two MSPs outside the  half-mass radius of \object{NGC~6752} which could have been ejected from the GC through interactions with a central massive object.

From early X-ray observations of GCs, we know that they are efficient at producing X-ray binaries in their core compared to the field. 
Using the eleven known bright LMXBs in GCs (now thirteen), \citet{VH87} showed that LMXBs in GCs are produced dynamically through exchange encounters of isolated neutron stars with primordial binaries as opposed to the much less probable evolution of a primordial binary into an LMXB.
Observations also support the fact that qLMXBs in GCs scale with the cluster encounter rate \citep{GBW03,Pooley+03}, implying that qLMXBs are formed through dynamical processes which occur in dense stellar systems.
Concerning CVs, following their discovery in significant numbers in \object{47~Tuc} \citep{GHEM01}, it has been pointed out following population synthesis studies \citep{Ivanova+06,Trenti+07} and from observations \citep[e.g.][]{Webb+04,PH06} that they may be formed in GCs either dynamically or, for a lower fraction, from primordial binaries.

Neglecting the role of binaries, we know that GC evolution leads to a core collapse followed by the GC disruption on a timescale shorter than the mean age of GCs, estimated to be $11.5\pm2.6$~Gyr \citep{CGCFP00}. This core collapse must have been delayed by an internal energy source to explain the GC longevity, and binaries could play this role \citep[see][ for a review]{Hut+92,Hut+03}.
In the core, binaries are subject to encounters and hard binaries become harder while transfering their energy to passing stars. This scenario leads to a global heating of the core, and even a small population of close binaries can drive the evolution of the entire cluster \citep{Hut+92}.
Some GCs could also contain an intermediate mass black hole (IMBH) of $\sim10^3\mathrm{~M_{\sun}}$ or more in their core to explain the distribution of stars in some clusters, and the stability on large time scales. 
The presence of an IMBH was claimed for \object{M~15} \citep[][ see also Ho~et~al. 2003]{Gerssen+02}, for G1 in the galaxy \object{M~31} \citep{GRH02,GRH05}, and for extragalactic GCs hosting an ultra luminous X-ray source \citep{Maccarone+07}.

The targets studied in this paper are two very different GCs. \object{NGC~2808} is a massive and concentrated cluster. \object{NGC~4372} on the contrary is less dense with fewer stars, being also a very low metallicity cluster. Their parameters are listed in Table~\ref{table:gcparam}.

\begin{table}
\begin{minipage}[t]{\columnwidth}
\caption{Globular cluster parameters from \citet[][ updated Feb. 2003]{Harris96}}             
\label{table:gcparam}      
\centering                          
\renewcommand{\footnoterule}{}  
\begin{tabular}{lcc}        
\hline\hline                 
Parameters & NGC~2808 & NGC~4372 \\    
\hline                        
Right Ascension (J2000)           & 9$^h$12$^m$02.6$^s$         & 12$^h$25$^m$45.4$^s$ \\
Declination (J2000)               & $-$64\degr51\arcmin47\arcsec\ & $-$72\degr39\arcmin33\arcsec\ \\
Distance [kpc]                    & 9.6                         & 5.8 \\
Core radius, $r_c$ [\arcmin]      & 0.26                        & 1.75 \\
Half mass radius, $r_h$ [\arcmin] & 0.76                        & 3.90 \\
Tidal radius, $r_t$ [\arcmin]     & 15.55                       & 34.82 \\
$r_c/r_h$                         & 0.34                        & 0.45 \\
Half-mass relaxation time [yr]    & $1.35\times10^{9}$          & $3.89\times10^{9}$ \\
Mass\footnote{calculated from the relation $M/M_{\sun}=3\times10^{0.4(M_v^{\sun}-M_v)}$ using the absolute visual magnitude.} [$M_{\sun}$]
                                  & $1.46\times10^{6}$          & $3.08\times10^{5}$ \\
Metallicity, [Fe/H]               & $-$1.15                     & $-$2.09 \\
\hline                        
\end{tabular}
\end{minipage}
\end{table}

\object{NGC~2808} has already been studied in X-rays with the GIS instrument on board ASCA observatory. Only one source was reported in the GIS catalogue \citep{Ueda+01} with a 1.35\arcmin\ error circle located at 7.5\arcmin\ from the cluster center.
No sources were reported during the ROSAT All Sky Survey observations \citep{Voges+99}. 
The core of \object{NGC~2808} was also observed with the Space Telescope Imaging Spectrograph (STIS) on board the Hubble Space Telescope (HST) in 2000 with ultra-violet (UV) filters F25QTZ (far-UV band centered at 159~nm) and F25CN270 (near-UV band centered at 270~nm). \citet{Dieball+05} looked for white dwarfs (WDs) and CVs in this data. They found $\sim$40~WD and $\sim$60~CV candidates in the field of view. Two of the CV candidates are variable (UV~222 and UV~397), and another has an optical counterpart (UV~170).
\object{NGC~2808} has also been observed in detail in the optical, and since \citet{Harris74}, it has been known that the horizontal branch in \object{NGC~2808} is unusual \citep[see also][]{Bedin+00,Carretta+06}. The main sequence is separated into three branches, possibly due to successive rounds of star formation, with different helium abundances \citep{Piotto+07}.

In the optical, the colour-magnitude diagram of \object{NGC~4372} indicates an old cluster ($15\pm4$~Gyr), with high reddening, but no special features \citep{Alcaino+91}. \citet{KK93} found 19 variable stars, of which one has a light curve consistent with an eclipsing CV of period 0.4~days.
In the X-ray, the ROSAT Observatory, with the HRI instrument, detected 19 sources \citep{JVH96}, of which 9 fall in our field of view. All these sources are located outside of the half-mass radius of the cluster, and none are consistent with the variable stars detected in the optical.


\section{X-ray observations and data processing}

\object{NGC~2808} was observed on February $1^{st}$ 2005, for 41.8~kiloseconds~(ks) with the three European Photon Imaging Cameras (EPIC MOS1, MOS2 and pn) on board the XMM-\textit{Newton} observatory, in imaging mode, using a full frame window and a medium filter. 

The observation of \object{NGC~4372} was performed under the same conditions on March 23--$24^{th}$ 2005, for 29.7~ks, but the MOS1~CCD6 was inoperative (micro-meteorite event on March $9^{th}$ 2005), and the pn data was lost due to technical issues.

\subsection{Data reduction and filtering}
\label{section:datared}

We processed the data using the XMM-\textit{Newton} Science Analysis System v7.0 (SAS). We used the \textit{emproc} and \textit{epproc} scripts with the most recent calibration data files to reduce the EPIC observation data files (ODFs). During this step, bad events mostly due to bad rows, edge effects, and cosmic ray events were flagged.
We filtered the three resulting event lists for event patterns in order to maximise the signal-to-noise ratio against non X-ray events. We selected only calibrated patterns, i.e. simple and double events for pn data and single to quadruple events for MOS data.
For pn events below 500~eV, we selected only single events because in this energy band non X-ray events also affect double events \citep[][ \S3.3.7]{XMM-UHB}.

Based on the light curve of single events exceeding 10~keV we identified periods of high background, due to soft proton flares, and selected good time intervals for the observation. For \object{NGC~2808}, this operation leads to 38.0, 37.3 and 30.2~ks of clean observation for MOS1, MOS2 and pn respectively. For \object{NGC~4372}, the observation was highly affected by flares and only 15.7~ks for MOS1 and 17.2~ks for MOS2 remain after filtering. We note that flaring activity is continuous during the exposure and events with energy above 2~keV are affected by a high noise, even after filtering.

\subsection{Source detection}

The list of events was divided into three energy bands (0.5--1.5, 1.5--3, and 3--10~keV) to allow us to derive spectral colours.
The source detection was done for all available data simultaneously (MOS1, MOS2 and pn where available).

We performed the source detection using the script \textit{edetect\_chain} which first calculates the live time, the vignetting, the sensitivity map and the background map for each detector and each energy band, and then calls a sliding box algorithm. 
Finally we ran the task \textit{emldetect}. For each source, the task performs a point spread function (PSF) fitting, for all the available detectors and for the three bands simultaneously, refines the coordinates, and gives the count rates, the hardness ratios, and the maximum likelihood (ML) of each source candidate. 

The fluxes were obtained by providing the energy conversion factors (ECFs, in units of $10^{11}\mathrm{~count~cm^{2}~erg^{-1}}$) which allow the direct conversion of count rates into fluxes (flux~=~rate~/~ECF).
These factors were calculated in each energy band and for each detector by extracting an on-axis source and generating detector response files for the source (using \textit{rmfgen} and \textit{arfgen} SAS tasks). These response files were used to create a fake spectrum corresponding to a common model: a power law spectrum with $\Gamma=1.7$ (mean spectrum of detected sources) and the absorption ($N_H$) of the cluster. 
For \object{NGC~2808}, the absorption of $1.2\times10^{21}\mathrm{~cm^{-2}}$ was calculated from the reddening of optical observations \citep{Bedin+00} with the relation computed by \citet{BH78}.
For \object{NGC~4372}, the reddening estimated by \citet{Alcaino+91} allows us to determine an absorption of $2.8\times10^{21}\mathrm{~cm^{-2}}$.
Finally, ECFs were calculated by dividing the count rates of the fake spectrum by the model fluxes. The obtained fluxes were then converted to unabsorbed fluxes.
This method gives a reliable estimation of the ECFs we need for our energy bands. To give an idea of the errors on these values, by changing the spectral index of the model to $\Gamma=2.0$, the fluxes changed by 1.5\%, 2\%, and 8\%, in the energy bands 0.5--1.5, 1.5--3 and 3--10~keV respectively.

We processed the entire field of view with a spatial binning factor of 80, giving images with square pixels of side~4\arcsec. This gives pixels of a similar size to the pn pixel size (${4.1\arcsec\times4.1\arcsec}$), but larger than the MOS pixels (${1.1\arcsec\times1.1\arcsec}$).
The binning is sufficiently small to sample the PSF of the two detectors correctly, where the pn and MOS PSF~FWHM (Full Width Half Maximum) are 6\arcsec\ and 5\arcsec\ respectively \citep[][ \S3.1]{XMM-UHB}.
We used a sliding box of $5\times5$ pixels to detect the sources, and selected sources with a ML greater than 10 ($4\sigma$ detection).

\begin{figure}
\centering
\includegraphics[width=\columnwidth]{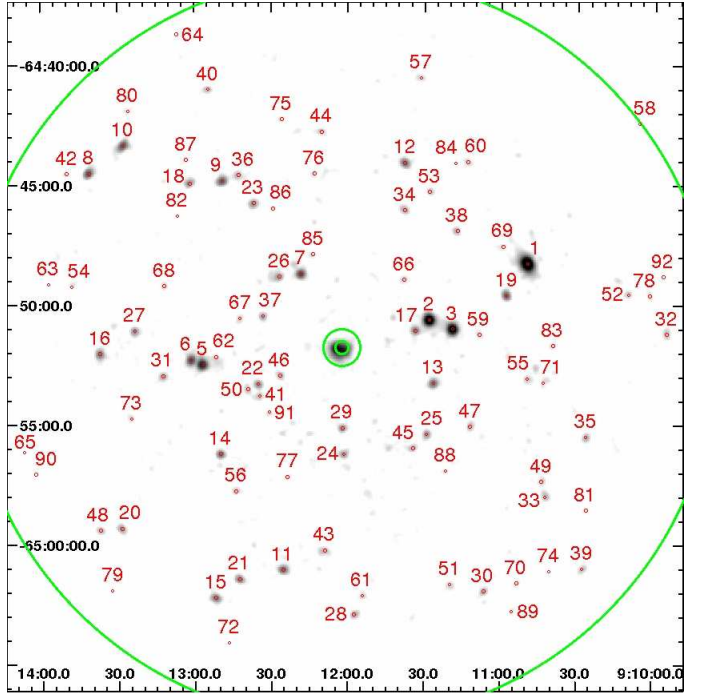}
\caption{\label{fig:2808map}
Combined image of the XMM-\textit{Newton} observation of \object{NGC~2808}. The three centered circles shown represent the core, half-mass and tidal radii. The detected sources are plotted with their 90\% error circles. 
We used a spatial binning factor of 80 and the image was smoothed with a Gaussian filter. For this reason the core appears blurry, a zoom is shown in Fig.~\ref{fig:2808}.}
\end{figure}

\begin{figure*}[!]
\centering
\includegraphics[width=16cm]{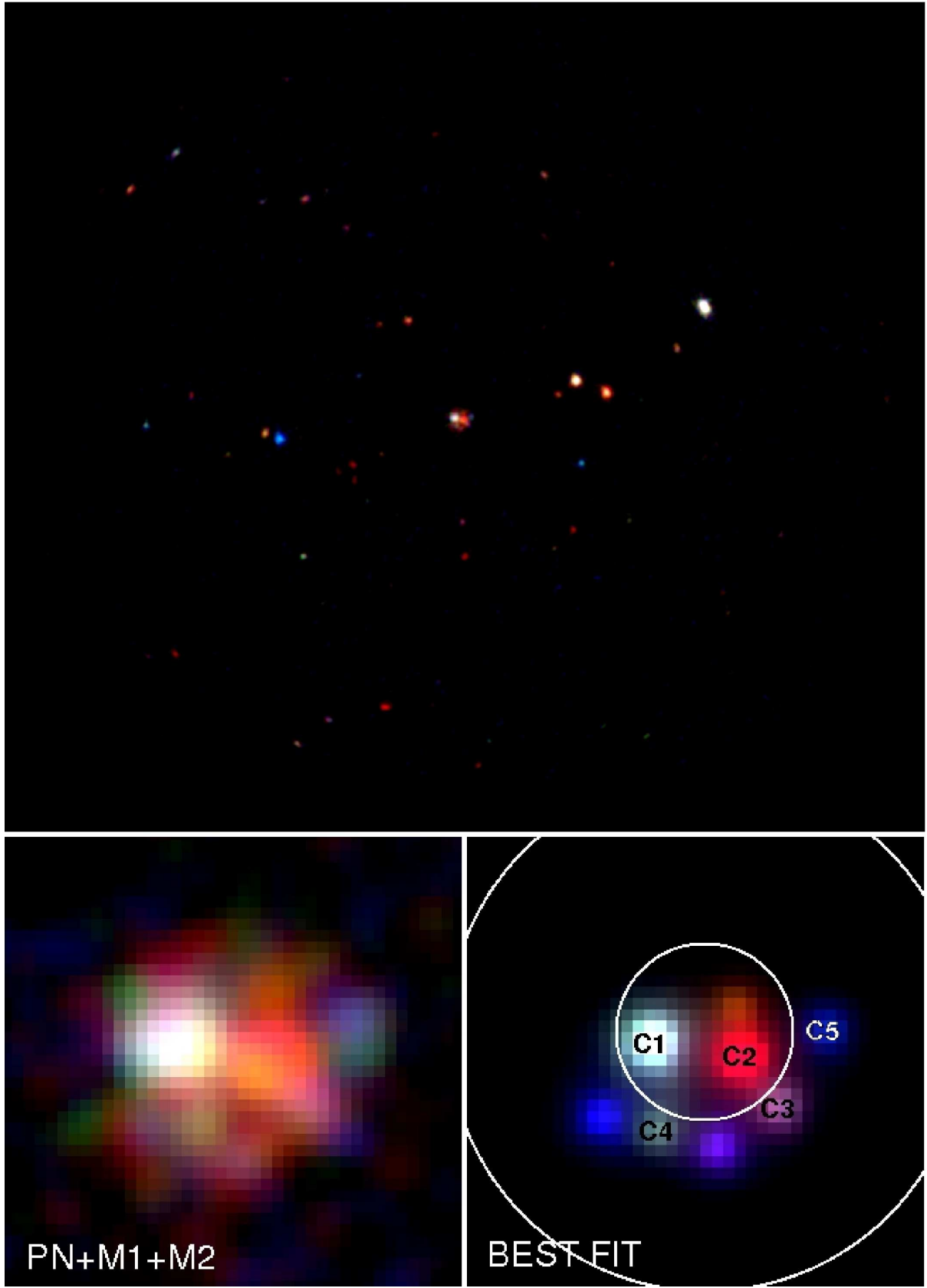}
\caption{\label{fig:2808}
\textit{Top}: XMM-\textit{Newton} observation of \object{NGC~2808}. Colours correspond to different energy bands, red:~0.5--1.5~keV, green:~1.5--3~keV, blue:~3--10~keV. The field of view is 30\arcmin\ across. The binning factor is 40 and the image was smoothed with a Gaussian filter. \textit{Bottom}:~A~zoomed combined image of the three detectors is shown on the left and the image on the right shows a reconstruction of the core with eight sources (only five are detected at $4\sigma$, the other three are detected at $2.5\sigma$). The core and the half-mass radii are shown.}
\end{figure*}

For \object{NGC~2808}, this leads to the detection of 92~sources, all of which were visually verified for each detector. Five sources fall outside the pn detector, four for MOS1 and three others for MOS2. In the central region of the detectors, along the line of sight of the center of the GC, we noted that a source has a complex form. In this region, the PSF is narrower and better defined and the vignetting is the lowest. 
We therefore reprocessed the data with a spatial binning of 40 to get squared pixels of side~2\arcsec\ and thus better sample the PSF for MOS detectors. We listed from visual inspection eight possible sources, and used the task \textit{emldetect} to simultaneously fit those source candidates.
We detect five additional sources at a minimum of $4\sigma$ (C1 to C5) in the half mass radius of the cluster, of which 3 fall inside the core radius. Three others are detected at $2.5\sigma$.
We have a total number of 96~sources detected above $4\sigma$. Their properties are listed in Table~\ref{table:2808src}, the combined image is presented in Fig.~\ref{fig:2808map}, and a colour image, with a zoom of the center, in Fig.~\ref{fig:2808}.

\addtocounter{table}{1}

\begin{figure}
\centering
\includegraphics[height=\columnwidth]{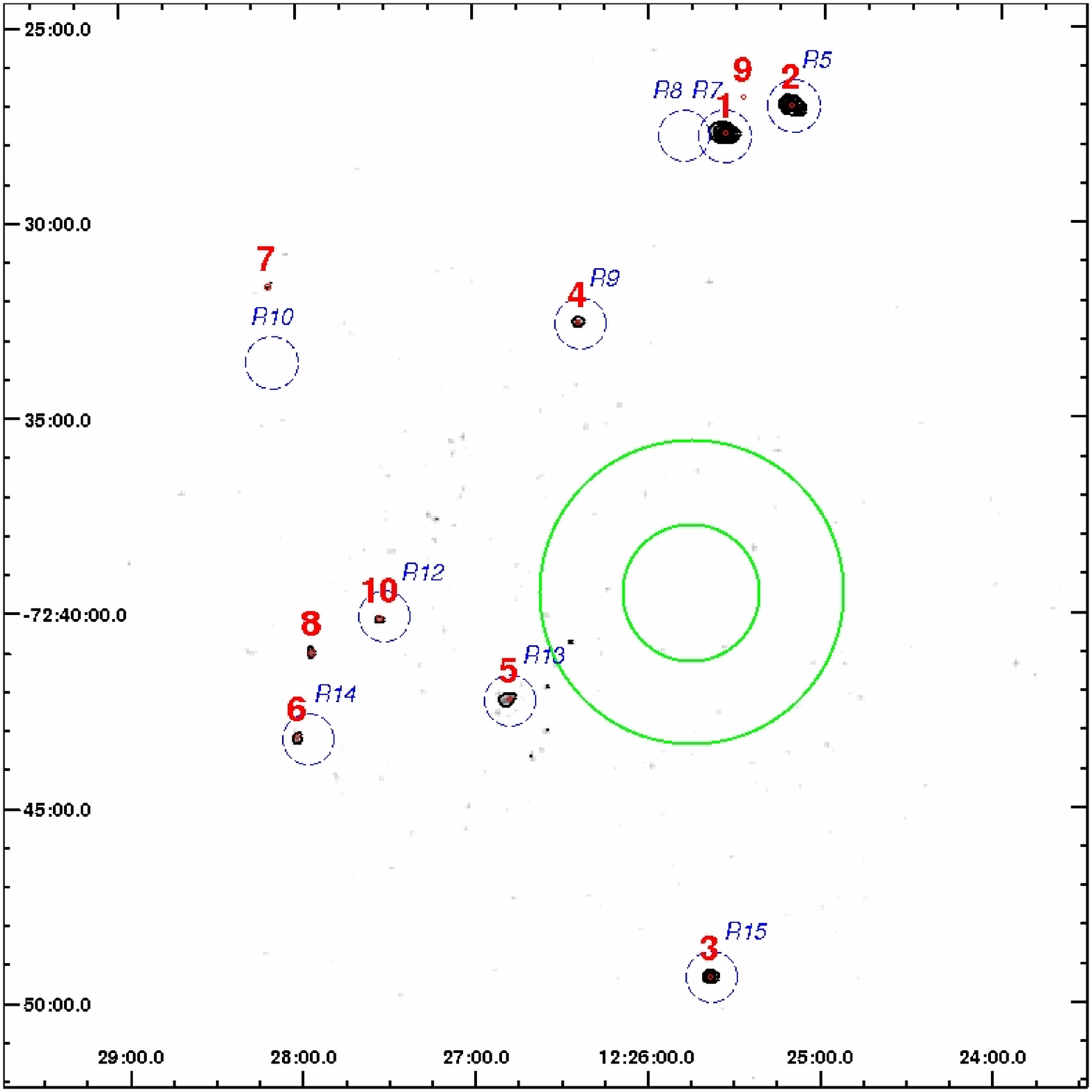}
\caption{\label{fig:4372}
Contour map of the XMM-\textit{Newton} MOS2 observation of \object{NGC~4372}. Core and half-mass radii are shown. The detected sources are plotted with their 90\% error circles and with contours at 3, 5 and $10\sigma$. Small circles are XMM-\textit{Newton} sources and bigger dashed circles are ROSAT sources with their error circle as reported by \citet{JVH96}.}
\end{figure}

For \object{NGC~4372}, 10~sources were detected. Three sources were detected with the MOS2 only, in the region where the MOS1~CCD6 fell. None of these sources are located inside the half mass radius. The X-ray source properties are listed in Table~\ref{table:4372src}, and the contour map is presented in Fig.~\ref{fig:4372}.

\addtocounter{table}{1}

\section{Ultra-violet observations and data reduction}

Along with the EPIC instruments, the Optical Monitor (OM) on board the XMM-\textit{Newton} observatory performed three exposures of 4\,000~seconds for both \object{NGC~2808} and \object{NGC~4372} with the UVM2 filter, centered at 231~nm in the UV band. For this filter, the PSF~FWHM is 1.8\arcsec\ and the field of view is approximatively $16\arcmin\times16\arcmin$.

The UV data was processed with the SAS task \textit{omichain}. This script removes bad pixels, performs spatial calibration, and source detection for each image. For each source, the instrumental magnitude is evaluated from the count rate. Finally the resulting images and source lists are merged. We considered a significance threshold of $3\sigma$ for detected sources.

\begin{table}
\caption{List of UV counterparts in the \object{NGC~2808} field of view. The X-ray source ID is given with the position of the UV source (90\% error is 2.0\arcsec), the offset position between UV and X-ray source, the offset from the cluster center, and the UVM2 magnitude.}             
\label{table:uv}      
\centering                          
\begin{tabular}{@{~}c@{~~~}c@{$^h$}c@{$^m$}c@{$^s$~~~}c@{\degr}c@{\arcmin}c@{\arcsec~~~}c@{~~}cc@{~~~}c@{~}}        
\hline\hline                 
ID & \multicolumn{3}{c}{RA$_{2000}$} & \multicolumn{3}{c}{Dec$_{2000}$} & Offset & Offset & UVM2 \\
   & \multicolumn{3}{c}{           } & \multicolumn{3}{c}{            } &  X-ray & center &      \\
\hline                        
 1 & 09&10&49.9 & $-$64&48&15.48 & 1.18\arcsec & 8.51\arcmin & $18.40 \pm 0.01$ \\
 2 & 09&11&28.3 & $-$64&50&36.60 & 1.61\arcsec & 3.85\arcmin & $18.32 \pm 0.01$ \\
17 & 09&11&33.6 & $-$64&51&04.32 & 0.54\arcsec & 3.17\arcmin & $10.95 \pm 0.01$ \\
22 & 09&12&35.3 & $-$64&53&19.32 & 2.06\arcsec & 3.77\arcmin & $18.01 \pm 0.01$ \\
24 & 09&12&01.7 & $-$64&56&12.84 & 1.69\arcsec & 4.46\arcmin & $14.80 \pm 0.01$ \\
29 & 09&12&02.2 & $-$64&55&10.56 & 2.54\arcsec & 3.35\arcmin & $17.04 \pm 0.01$ \\
50 & 09&12&02.2 & $-$64&55&10.56 & 2.52\arcsec & 4.27\arcmin & $18.97 \pm 0.01$ \\
77 & 09&12&23.5 & $-$64&57&11.88 & 1.63\arcsec & 5.84\arcmin & $18.05 \pm 0.01$ \\
85 & 09&12&13.7 & $-$64&47&52.80 & 1.24\arcsec & 4.07\arcmin & $17.24 \pm 0.01$ \\
\hline                        
\end{tabular}
\end{table}

For \object{NGC~2808}, we detected 598 sources at a limiting UVM2 magnitude of 19.3. The region inside the half-mass radius is overcrowded and poorly resolved. Approximatively 45 X-ray sources fall inside the OM field of view, for which we found nine matching UV sources. 
The properties of the UV counterparts are listed in Table~\ref{table:uv}.
We took into account the X-ray position accuracy of the sources ($\sim3.6\arcsec$, 90\% error circles) and the UV position accuracy ($\sim2.0\arcsec$, 90\% error circles). We thus kept matching UV sources that have an offset with the X-ray source lower than 5.6\arcsec. The X-ray sources with UV counterparts are all located outside the half-mass radius, and are therefore probably background or foreground sources.

We performed a Monte-Carlo simulation to evaluate the probability that the superposition of an X-ray source and a UV source occured by chance.
We excluded the central region inside a radius of 2.4\arcmin\ where the UV sources are not resolved. In the remaining area the X-ray and UV sources appeared to be uniformly distributed. The simulation leads to $1.1\pm1.1$ sources aligned fortuitously. We conclude that $8\pm1$ of the UV counterparts could be associated with their corresponding X-ray source with a probability of 99.9927\%.

For \object{NGC~4372}, 272 sources were detected at a limiting UVM2 magnitude of 19.6. A bright A0 star in the field causes an out-of-focus ghost image (smoke ring) and lead to $\sim10$ spurious detections. Only three X-ray sources fall in the region observed by the OM, and none has a UV counterpart.


\section{The X-ray sources in NGC~2808}

\subsection{Members of NGC~2808}
\label{section:members}

Most of the sources detected are background or foreground sources, but some of them are members of the cluster. If we assume no cosmic variance in the distribution of sources in the sky \citep[see][]{Yang+03}, we can compare our observation with one without a GC to determine statistically the number of sources belonging to the cluster. We evaluated the number of background sources based on observations of the \object{Lockman~Hole} (LH) with XMM-\textit{Newton}. In this field, centered on the sky position RA$_{2000}$~10$^h$52$^m$43$^s$, Dec$_{2000}$-$+$57\degr28\arcmin48\arcsec, the absorption is very low, $N_H=5.7\times10^{19}\mathrm{~cm^{-2}}$ \citep{LJM86}.

We divided the XMM-\textit{Newton} field of view into several annuli to take into account the vignetting, which becomes more important towards the edge of the field of view. The annuli are centered on the center of the GC, which is also the center of the sensitivity map of the detectors, and their size was chosen to encircle at least 5 detected sources, and when possible 20 sources.
For each annulus, we evaluated the background count rate in several 15\arcsec\ radius regions without sources assuming a correction for the vignetting. A minimum detectable count rate was estimated to be equivalent to a ML of 10 above this background count rate. We assumed a power law spectrum with $\Gamma=1.7$ with the absorption of the field of view, and entered the count rates into the tool WebPIMMS\footnote{http://heasarc.gsfc.nasa.gov/Tools/w3pimms.html}~v3.9b \citep{Mukai93} to determine the minimum detectable unabsorbed flux of each annulus. In the center, this limiting flux is ${F_{\mathrm{0.5-10~keV}}=4.1\times10^{-15}\mathrm{~erg~cm^{-2}~s^{-1}}}$, and goes up to ${6.1\times10^{-15}\mathrm{~erg~cm^{-2}~s^{-1}}}$ in the last annulus. This lead to a limiting luminosity of ${L_{\mathrm{0.5-10~keV}}=4.5\times10^{31}\mathrm{~erg~s^{-1}}}$ for sources in the core of \object{NGC~2808}.

We used the $log(N)-log(S)$ relation derived by \citet{HMS05} from a deep observation of the LH in the energy band 0.5--2.0~keV. Our estimated minimum detectable flux allows us to determine the number of sources expected per square degree. We took into account the quadratic sum of two errors: an error of 10\% on the flux, and an imprecision of 10\% on the $log(N)-log(S)$ relation. Results are reported in Table~\ref{table:lnls}.

\begin{figure}
\centering
\includegraphics[width=\columnwidth]{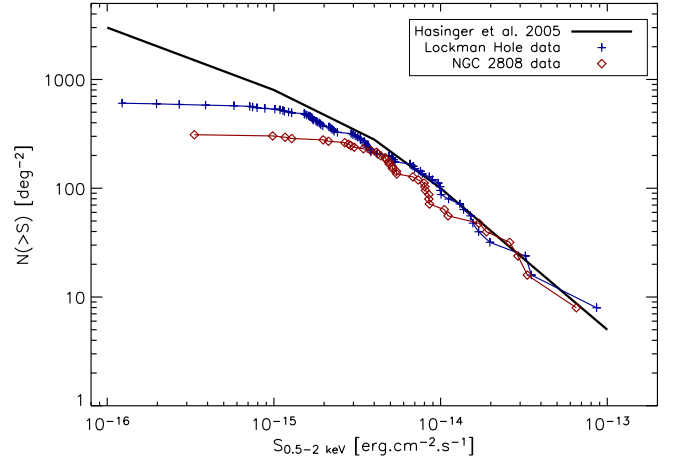}
\caption{\label{fig:lnls}
$Log(N)-log(S)$ diagram in the band 0.5--2 keV. S is the limiting flux and N the number of sources. The empirical relation of \citet{HMS05} is reported for comparison.}
\end{figure}

As a consistency check, we processed a long observation of 90~ks of the LH that took place on November $27^{th}$ 2002. This observation used a full frame window and a medium filter as for our observations. We performed exactly the same data processing that we used for our observations, and plotted the $log(N)-log(S)$ relation for the energy band 0.5--2~keV (see Fig.~\ref{fig:lnls}). We also plotted the $log(N)-log(S)$ relation using the sources detected in the same band with a correction for the higher absorption in the field of view of \object{NGC~2808}. The shape of the curves are similar, and we can see the flux detection limit in the different fields due to exposure time. The $log(N)-log(S)$ relation is therefore applicable to our data.

The results reported in Table~\ref{table:lnls} indicate that the five sources located in the center of the field of view are likely to be related to the cluster. The probability of membership if we assume a Poisson distribution is 99.9985\%.
Moreover a possible excess of sources appears in the annuli between radii 3.6\arcmin~and~6.5\arcmin\ ($4.5\mathrm{~r_{h}}$~to~$8\mathrm{~r_{h}}$). The probability of membership to the cluster is 96.75\%. As we are dealing with low numbers, statistical fluctuations might explain this excess, but we have nonetheless included these sources in our analysis.

\begin{table}
\caption{Expected and detected sources for \object{NGC~2808} field of view.}             
\label{table:lnls}      
\centering                          
\begin{tabular}{r@{ -- }lccc}        
\hline\hline                 
\multicolumn{2}{c}{Annulus}  & Expected & Detected \\    
\multicolumn{2}{c}{(\arcmin)} & 0.5--2 keV & 0.5--2 keV \\    
\hline                        
   0    & 0.76 &  0.30 $\pm$ 0.06 &  5  \\
   0.76 & 3.6  &  6.27 $\pm$ 1.24 &  5  \\
   3.6  & 6.5  & 14.31 $\pm$ 2.94 & 20  \\
   6.5  & 9    & 17.58 $\pm$ 3.79 & 20  \\ 
   9    & 12   & 21.99 $\pm$ 5.88 & 20  \\ 
\hline                        
\end{tabular}
\end{table}

\subsection{Spectral analysis}
\label{section:sp}

\begin{figure}
\centering
\includegraphics[width=\columnwidth]{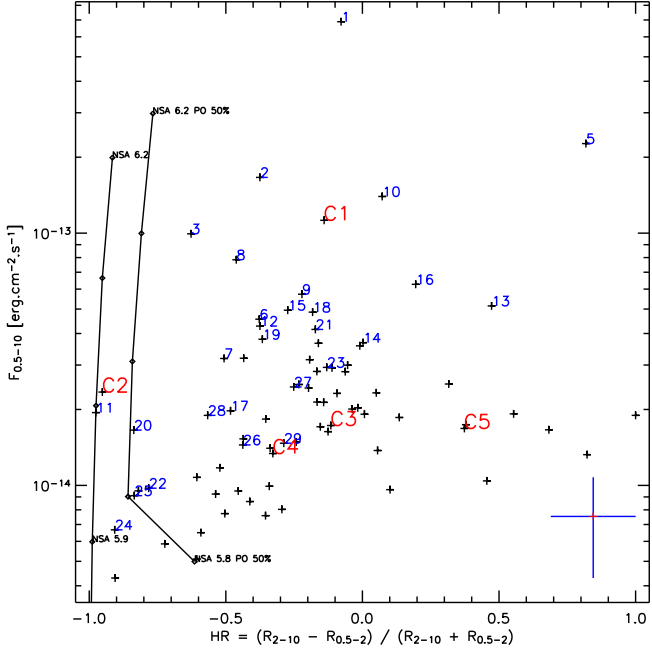}
\caption{\label{fig:2808hrf}
Flux-colour diagram of \object{NGC~2808} sources. For clarity, only the 30 brightest sources are enumerated, the flux values for all the sources are listed in Table~\ref{table:2808src}. $R$ is the count rate for the given energy band in keV. A typical error bar is shown at the bottom right. Black lines represent the following models, assuming an absorption of $1.2\times10^{21}\mathrm{~cm^{-2}}$: \newline \textbf{NSA}:~neutron star with hydrogen atmosphere, mass $1.4\mathrm{M_{\sun}}$, radius 12~km, distance of the cluster and $log(T_\mathrm{eff})=5.9$, 6, 6.1, 6.2. \newline \textbf{NSA PO}:~NSA and a power law with photon index of 1 and contribuing 50\% of the flux.}
\end{figure}

\begin{figure}
\centering
\includegraphics[width=\columnwidth]{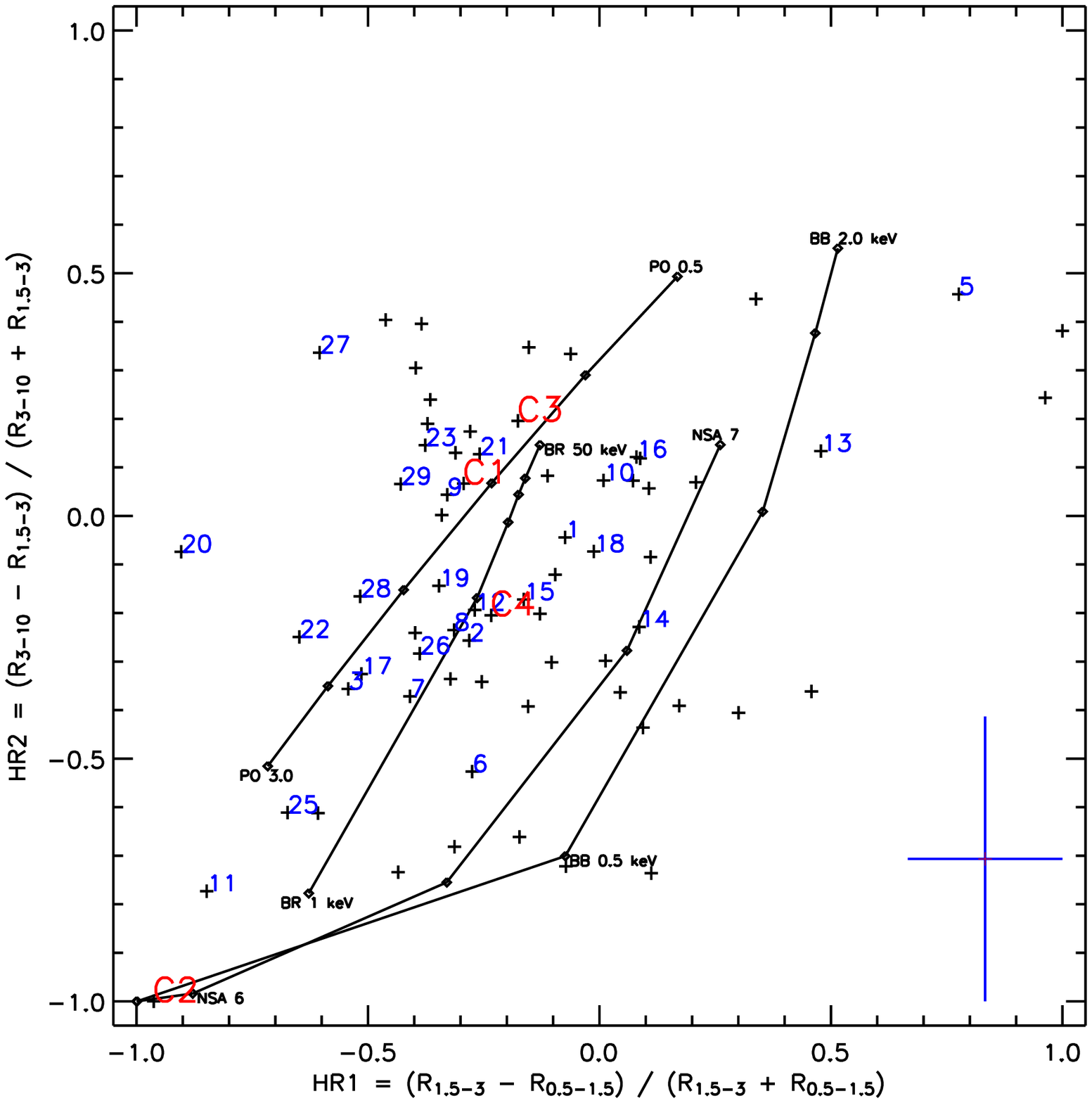}
\caption{\label{fig:2808hr}
Colour-colour diagram of \object{NGC~2808} sources. Same comments as for Fig.~\ref{fig:2808hrf}.
\newline \textbf{PO}:~power law with photon indices 3, 2.5, 2, 1.5, 1, 0.5. \newline \textbf{BR}:~thermal bremsstrahlung with temperatures 1, 5, 10, 15, 20, 50~keV. \newline \textbf{BB}:~blackbody spectrum with temperatures 0.1, 0.5, 1, 1.5, 2~keV. \newline \textbf{NSA}:~neutron star with hydrogen atmosphere, mass $1.4\mathrm{M_{\sun}}$, radius 12~km, distance of the cluster and $log(T_\mathrm{eff})=5$, 6, 6.5, 6.8, 7.}
\end{figure}

We first plotted the hardness ratios obtained during the source detection for sources with more than 50~counts and more than 3~counts in each energy band (Figs.~\ref{fig:2808hrf} and~\ref{fig:2808hr}). We show the tracks of some spectral models. In Fig.~\ref{fig:2808hrf}, possible qLMXBs are located on the left side of the diagram, while very absorbed sources are seen on the right side of the diagram. The most luminous and absorbed sources have colours that are similar to these of extragalactic objects as discussed in Sect.~\ref{section:discuss}. In Fig.~\ref{fig:2808hr}, the qLMXB candidates should be located at the bottom left, and CVs in the middle of the diagram around the power law branch with photon indices 1 to 1.5 and the bremsstrahlung branch with temperatures 10 to 50~keV mbox{\citep[e.g.][]{Richman96,BWO05}}.

For the brightest sources, we extracted and fitted the spectra. We used an extraction radius of 30\arcsec\ whenever possible, so 80\% of the encircled energy was included. In the crowded region the extraction radius was reduced to 8\arcsec, corresponding to 50\% encircled energy \citep[][ \S3.2.1]{XMM-UHB}. A correction is taken into account in the instrumental response files. We extracted a background with the same extraction radius for each source in a region without sources on the same CCD, with the same vignetting, and when possible at the same distance from CCD readout node for pn CCDs.

We used the task \textit{evselect} with a binning of 15~eV for MOS data and 5~eV for pn data as recommended \citep[][ \S4.9.1]{SAS-USG}. Then for each spectrum we used \textit{rmfgen} and \textit{arfgen} to generate the instrumental response files for a point source, i.e. the redistribution matrix file (RMF) and the ancillary response file (ARF).
We fitted the data with Xspec~v11.3.2 \citep{Arnaud96}. A binning greater than 20~counts allowed the use of the $\chi^2$ minimization criterion, but when there were insufficient counts per bin (under 20 per bin), we used the Cash statistic \citep{Cash79} which provides a goodness-of-fit criterion similar to that of $\chi^2$.

\begin{figure}
\centering
\includegraphics[width=\columnwidth]{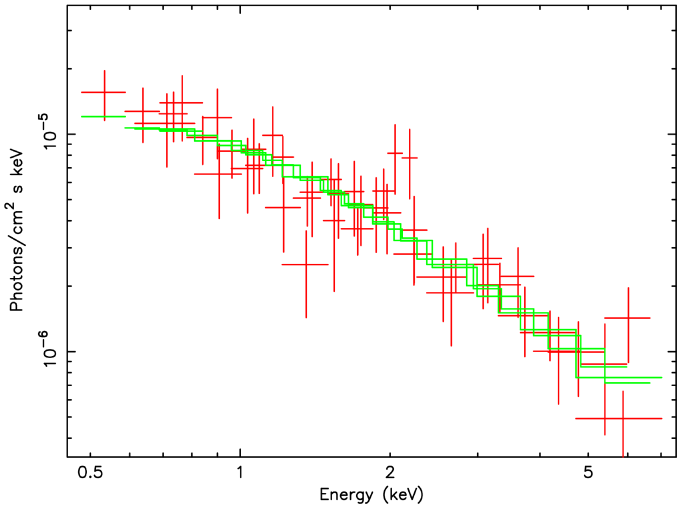}
\caption{\label{fig:2808_C1_sp}
Spectrum of C1 (\object{NGC~2808}) fitted with a power law model and the absorption of the cluster.}
\end{figure}

\begin{figure}
\centering
\includegraphics[width=\columnwidth]{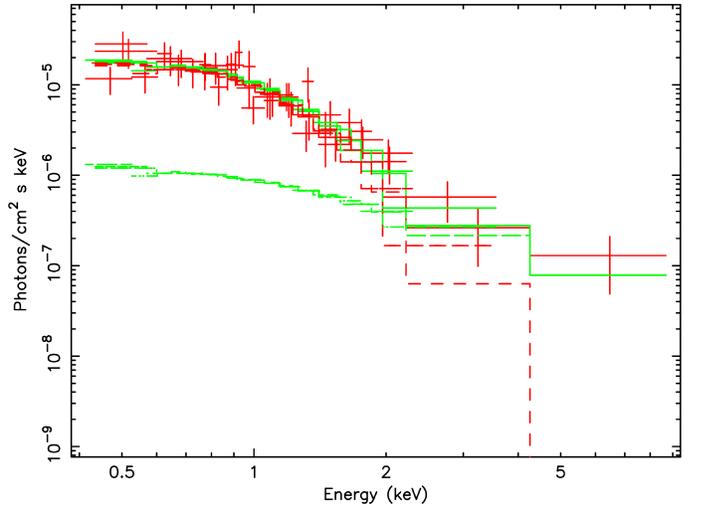}
\caption{\label{fig:2808_C2_sp}
Spectrum of C2 (\object{NGC~2808}) fitted with a NSA model and the absorption of the cluster. The contribution of C1 also appears as a power law.}
\end{figure}

We tried simple models included in Xspec such as a power law, a bremsstrahlung, a black body, a Raymond Smith or a mekal fit. For very soft sources we tried a hydrogen atmosphere model \citep{ZPS96}, assuming the distance of the GC, a mass of $1.4\mathrm{~M_{\sun}}$ and a radius of 12~km for a neutron star. These parameters correspond to other neutron stars detected previously in GCs \citep[][]{WB07,HGLE03}.
When it was clear that a simple model was insufficient to fit the data we tried composite models.
The results of the spectral fitting are provided in Table~\ref{table:2808fit} and are discussed in Sect.~\ref{section:discuss}. 
We also provide the spectra of C1 and C2 (Figs.~\ref{fig:2808_C1_sp}~and~\ref{fig:2808_C2_sp}).

\begin{table*}
\begin{minipage}[t]{\linewidth}
\caption{\label{table:2808fit}
Best fitting models to the spectra of sources in the \object{NGC~2808} field of view. The unabsorbed flux is in the 0.5--10 keV range [$\times$~10$^{-14}$~erg~cm$^{-2}$~s$^{-1}$] and for possible cluster sources the luminosity is given in the 0.5--10 keV range [$\times$~10$^{32}$~erg~s$^{-1}$]. The models tried are a power law (PO), an absorbed power law (APL), a bremsstrahlung (BR), a black body (BB), a Raymond-Smith (RS), a mekal (MK) and a neutron star with hydrogen atmosphere (NSA). We used models with one or two components (Comp. 1 and Comp. 2). The absorption $N_H~gal$ [$\times$~10$^{21}$~cm$^{-2}$] is frozen to the value of the cluster. Parameter values are photon-index $\Gamma$ of PO model, temperature $kT$ [keV] of BB, RS and MK models, and $log(T_\mathrm{eff})$ [K] of NSA model. $\chi^2$ and $C$ give the goodness of the fit, reported with the number of degrees of freedom ($dof$).}
\centering
\begin{tabular}{@{~}c@{~~}c@{~~}c@{~~}|@{~~}c@{~~}c@{~~}c@{~~}c@{~~}c@{~~}|@{~~}c@{~~}c@{~~}c|c@{~}c@{~}c@{~}}
\hline\hline       
Src & Flux & $N_{H~gal}$ & Model   & $N_{H}$ & $\Gamma$ & $kT$ & $log(T_\mathrm{eff})$ & Model   & $\Gamma$ & $kT$ &  $\chi^2$ & $C$ & $dof$ \\ 
ID  & (Lum) &            & Comp. 1 &         &          &      &                & Comp. 2 &          &      &                  &     &       \\ 
\hline                    
C1 & $8.6\pm1.5$ & 1.2 & PO & -- & $1.56\pm0.14$ & -- & -- & -- & -- & -- & -- & 37.54 & 43  \\
   & ($9.5\pm1.7$)       & 1.2 & BR & -- & -- & 12.28$^{+19.42}_{-5.03}$ & -- & -- & -- & -- & -- & 37.35 & 43  \\
\hline                    
C2 & $2.4\pm0.3$ & 1.2 & PO & -- & $2.8\pm0.2$ & -- & -- & -- & -- & -- & -- & 51.37 & 43  \\
   & ($2.6\pm0.4$)     & 1.2 & NSA & $0.98\pm0.04$ & -- & -- & $6.016\pm0.017$ & -- & -- & -- & -- & 56.45 & 43  \\
   &      & 1.2 & NSA & $0.82\pm0.40$ & -- & -- & $5.975\pm0.027$ & PO & 1.56 & -- & -- & 38.88 & 38  \\
\hline                    
 1 & $52.8\pm8.6$ & 1.2& PO  & -- & $1.31\pm0.05$ & -- & -- & -- & -- & -- & 143.38 & -- & 94  \\
   &              & 1.2 & APL
                       & $4.3\pm1.8$ & $1.7\pm0.2$ & -- & -- & PO & $4.7\pm1.4$ & -- & 98.07 & -- & 93  \\
\hline                    
 2 & $15.4\pm1.2$ & 1.2 & PO & -- & $1.9\pm0.1$ & -- & -- & -- & -- & -- & 48.28 & -- & 36  \\
\hline                    
 3 & $9.1\pm1.1$  & 1.2 & PO & -- & $2.6\pm0.1$ & -- & -- & -- & -- & -- & 46.36 & -- & 36  \\
\hline                    
 5 & 17.3$\pm$2.7 & 1.2 & APL & $35.03^{+18.63}_{-13.71}$ & $1.65\pm0.60$ & -- & -- & -- & -- & -- & 27.17 & -- & 20  \\
   &             & 1.2 & APL & 49.46$^{+23.53}_{-10.57}$ & 1.918$^{+0.89}_{-0.53}$ & -- & -- & PO & 2.0 & -- & 15.14 & -- & 19   \\
\hline                    
 13 & $6.4\pm0.$7 & 1.2 & PO & -- & $0.32\pm0.25$ & -- & -- & -- & -- & -- & -- & 53.95 & 38   \\
   &              & 1.2 & APL & $5.64^{+8.9}_{-4.93}$ & $0.75^{+0.64}_{-0.42}$ & -- & -- & PO & $9.5^{+0.5}_{-0.6}$ & -- & -- & 41.20 & 35   \\
\hline                    
 17 & $2.1\pm0.$6 & 1.2 & PO & -- & $2.1\pm0.3$ & -- & -- & -- & -- & -- & -- & 28.95 & 27  \\
    &             & 1.2 & MK & -- & -- & $2.7\pm1.0$ & -- & -- & -- & -- & -- & 30.63 & 27  \\
\hline                    
 22 & $1.1\pm0.6$   & 1.2 & BB & -- & -- & $0.16\pm0.02$ & -- & MK & -- & $76^{+4}_{-59}$ & -- & 49.54 & 40  \\
    & ($1.2\pm0.6$) &     &    &    &    &               &    &    &    &                 &    &       &     \\
\hline                    
 24 & $0.9\pm0.2$   & 1.2 & RS & -- & -- & $0.40\pm0.09$ & -- & -- & -- & -- & -- & 29.62 & 27  \\
    & ($1.0\pm0.2$) &     &    &    &    &               &    &    &    &    &    &       &     \\
\hline                    
 25 & $0.9\pm0.2$   & 1.2 & BB & -- & -- & $0.17\pm0.09$ & -- & MK & -- & $6.19\pm6.1$ & -- & 18.71 & 25  \\
    & ($1.0\pm0.2$) &     &    &    &    &               &    &    &    &              &    &       &     \\
\hline                    
\end{tabular}
\end{minipage}
\end{table*}

A global background spectrum was extracted for each EPIC camera in order to locate possible features in the background that could be present in the source spectra. We selected all regions without sources by excluding areas of 1\arcmin\ radius around detected sources. We produced the instrumental response files for a flat field.
Based on this spectrum, we fixed the limits of the energy bands we used: 0.4--15~keV for pn events, and 0.2--10~keV for MOS events. For all sources we could evaluate the flux in the 0.5--10~keV energy band.
For MOS, two strong lines are observed at 1.5 and 1.75~keV (Al~K$\alpha$ and Si~K$\alpha$ lines). For pn, the background spectrum shows a strong line feature around 1.5~keV (Al~K$\alpha$ line), fainter lines at 8 and 8.6~keV (Cu lines), and some features at 9.7 and 11~keV \citep[as seen in][ \S3.3.7]{XMM-UHB}. 

\subsection{Variability analysis}

\begin{figure}
\centering
\includegraphics[width=\columnwidth]{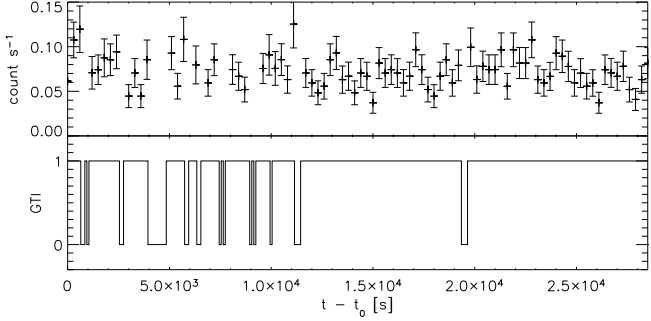}
\caption{\label{fig:2808_1_lc}
Lightcurve of source 1 (\object{NGC~2808}) and corresponding good time intervals (GTI) as defined in Sect.~\ref{section:datared}. Start time~$t_0$ is February $1^{st}$ 2005 4:57:28 (MJD 2453402.70657).}
\end{figure}

We performed variability analysis based on the pn data for sources with more than 300 pn counts, and for sources with a fitted spectrum and more than 100 pn counts. We extracted lightcurves for sources and backgrounds (same regions as for spectral analysis), adjusting the binning for each source to obtain a mean of 20 counts per bin after correction. We removed the first 12~ks of the observation as this part is affected by flares.
We corrected the pn events for losses due to e.g. vignetting or filters with the SAS task \textit{lccorr}. The time variability is examined with a Kolmogorov-Smirnov test using IDL/Astrolib\footnote{http://idlastro.gsfc.nasa.gov/} procedure \textit{kstwo} by comparing the source lightcurve to the background lightcurve. 
We also fitted the background subtracted lightcurve with a constant value, and calculated the $\chi^2$ of the fit. The results are reported in Table~\ref{table:2808fitvar}. Only source 1 seems to be variable with a significant probability. The lightcurve is presented in Fig.~\ref{fig:2808_1_lc} and discussed in Sect.~\ref{section:other2808}.

\begin{table}
\caption{\label{table:2808fitvar}      
Variability analysis of \object{NGC~2808} sources. We give for each source the Bin size [s], the Kolmogorov-Smirnov probability of the source lightcurve being differently distributed than the background lightcurve (K-S), and the $\chi^2$ with the number of degrees of freedom ($dof$) for the background subtracted lightcurve fitted with a constant.}
\centering                          
\begin{tabular}{ccccccc}        
\hline\hline                 
ID & Bin size & K-S & $\chi^2$ & $dof$ \\    
\hline                        
 C1 & 2400 & 0.15 & 6.08 & 10 \\
 C2 & 3000 & 0.31 & 20.05 & 9 \\
  1 &  300 & $1.05\times10^{-4}$ & 120.90 & 82 \\
  2 &  600 & 0.23 & 48.71 & 39 \\
  3 &  600 & 0.04 & 67.87 & 39 \\
  5 & 1200 & 0.33 & 31.90 & 21 \\
 13 & 2000 & 0.86 & 23.65 & 13 \\
 17 & 2400 & 0.15 & 31.08 & 10 \\
\hline                        
\end{tabular}
\end{table}

\section{The X-ray sources in NGC~4372}

\citet{JVH96} found 9 sources in \object{NGC~4372} with the ROSAT X-ray observatory in an equivalent region of the sky covered by our XMM-\textit{Newton} observation. Two of these sources were not detected in our observations (R8 and R10), and we detect three additional sources (7, 8 and 9). Three merged sources in the ROSAT image (R5, R7 and R8) are clearly resolved into two bright sources (1 and 2) and one faint and more diffuse source (9), see Fig.~\ref{fig:4372}.

We could not use a $log(N)-log(S)$ relation to discuss the distribution of sources, as in Sect.~\ref{section:members} for \object{NGC~2808}, because of the low number of sources detected which leads to large error bars. 
Moreover the high background noise due to high flaring activity does not allow us to estimate accurately the flux detection limit. 
From the faintest sources detected, this limit is around ${F_{\mathrm{0.5-10~keV}}\sim3\pm1\times~10^{-14}\mathrm{~erg~cm^{-2}~s^{-1}}}$, which leads to a limiting luminosity of ${L_{\mathrm{0.5-10~keV}}\sim10^{32}\mathrm{~erg~s^{-1}}}$ for a source in the core of the cluster.
From the count rate of the faintest sources detected by ROSAT in this region, their limiting flux (obtained with WebPIMMS) is ${F_{\mathrm{0.5-10~keV}}\sim5\pm2\times10^{-14}\mathrm{~erg~cm^{-2}~s^{-1}}}$, so we expect to detect all ROSAT sources in this region.
We note that our flux detection limit is comparable to the ROSAT limit, due to large cuts after flare filtering and the absence of pn data, but we increase the angular resolution and clearly resolve the sources. 

We performed the same analysis as in Sect.~\ref{section:sp}. 
We carried out spectral analysis for two sources with sufficient counts, in the same manner as described in Sect.~\ref{section:sp}. The results are reported in Table~\ref{table:4372fit}.

\begin{figure}
\centering
\includegraphics[width=\columnwidth]{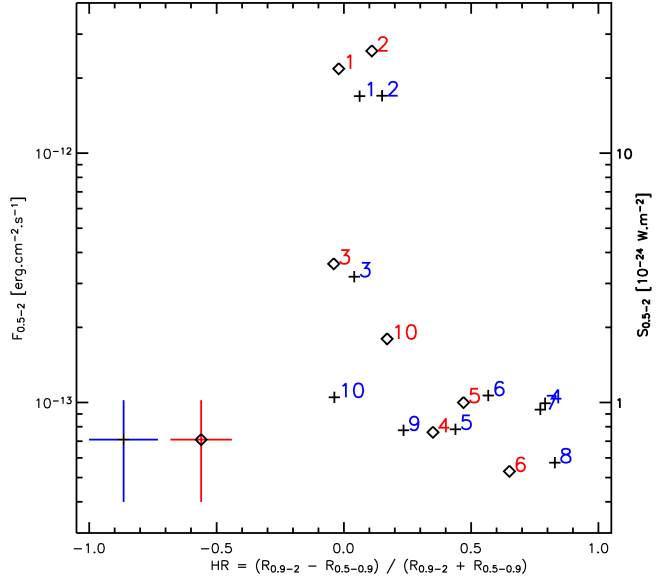}
\caption{\label{fig:4372hrf}
Flux-colour diagram of \object{NGC~4372} sources detected with XMM-\textit{Newton} (crosses, units on the left axis) and ROSAT (diamonds, units on the right axis). $R$ is the count rate for the given energy band in keV. Typical error bars are shown at the bottom left.}
\end{figure}

We compared hardness ratios given by \citet{JVH96} to the ones we obtained using the same energy bands. We plotted for both ROSAT and XMM-\textit{Newton} data a flux-colour diagram (Fig.~\ref{fig:4372hrf}). The two distributions are well correlated within the error bars, especially for bright sources, but we note that source 4 seems harder in our observation.

\begin{table}
\begin{minipage}[t]{\linewidth}
\caption{\label{table:4372fit}      
Best fitting models to spectra of sources in the \object{NGC~4372} field of view. Same as Table~\ref{table:2808fit}.}
\centering
\begin{tabular}{@{~}c@{~~}c@{~~}c@{~~}|@{~~}c@{~~}c@{~~}c@{~~}|@{~~}c@{~~}c@{~}}
\hline\hline       
Src & Flux &   $N_{H~gal}$ & Model & $\Gamma$ & $kT$ & $\chi^2$ & $dof$ \\ 
\hline                    
1 & $148\pm10$ & 2.8 & PO & $4.1\pm0.2$ & -- & 155.94 & 113  \\
  &            & 2.8 & BB & -- & $0.14\pm0.01$ & 153.70 & 113  \\
\hline                    
2 & $147\pm10$ & 2.8 & PO & $4.1\pm0.2$ & -- & 100.40 & 79  \\
  &            & 2.8 & BB & -- & $0.14\pm0.01$ & 99.02 & 79  \\
\hline                    
\end{tabular}
\end{minipage}
\end{table}


\section{Discussion}
\label{section:discuss}

We discuss here the presence of objects expected in globular clusters such as qLMXBs, CVs, MSPs and ABs, adding notes on individual sources. We also discuss some background and foreground sources which present unusual features.

\subsection{Low mass X-ray binaries in quiescence}

For 18 GCs that have been observed deep enough to detect all qLMXBs, we plotted the number of qLMXBs against the approximate encounter rate for a virialized system ($\rho_0^{1.5}r_c^2$), as done in \citet{GBW03}.
According to this correlation, one would expect $3\pm1$~qLMXBs in \object{NGC~2808}, and none in \object{NGC~4372}.

For \object{NGC~2808} we detect one source consistent with a qLMXB (C2, described below).
For \object{NGC~4372}, a qLMXB having the minimum luminosity of the 21 qLMXBs reported by \citet{Heinke+03} would have been detected. The lack of detection is therefore consistent with the prediction of zero qLMXB in this cluster.

\paragraph{NGC~2808 -- C2.}
This source is located in the core radius of \object{NGC~2808}. It is close to a bright source (C1, 15\arcsec) and the PSF wings are merged. From \citet[ \S3.2.1]{XMM-UHB}, we estimated that 10\% of the emission of C1 was present in the C2 spectrum. We fitted a model composed of 10\% of the C1 spectrum and a hydrogen atmosphere model for a neutron star. The parameters are given in Table~\ref{table:2808fit} and the spectrum in Fig.~\ref{fig:2808_C2_sp}. The hard emission in this spectrum appears to be due to C1.
We note that the unabsorbed luminosity ($2.6\pm0.4\times10^{32}\mathrm{~erg~s^{-1}}$ if it belongs to the cluster), and the X-ray spectrum of C2 are consistent with the qLMXB hypothesis. 

\subsection{Cataclysmic variables}

\citet{Ivanova+06}, using numerical simulations and taking into account different CV formation channels, predict about 200~CVs for a GC similar to \object{NGC~2808}. 
Of these, only a fraction can be detected in our data. 
From the empirical function of \citet{PH06} derived from fitting the number of bright CVs (at luminosities above ${L_{\mathrm{0.5-10~keV}}=4.25\times10^{31}\mathrm{~erg~s^{-1}}}$) against the specific encounter frequency, we estimate a population of $20^{+20}_{-10}$ bright CVs in \object{NGC~2808}.
Given the X-ray unabsorbed luminosity of the core (excluding C2) of ${L_{\mathrm{0.5-10~keV}}=1.4\times10^{33}\mathrm{~erg~s^{-1}}}$, we cannot expect more than 30 bright CVs in this region.

We detect 4 CV candidates in \object{NGC~2808} with luminosities down to ${L_{\mathrm{0.5-10~keV}}\sim3.0\times10^{32}\mathrm{~erg~s^{-1}}}$. A hidden unresolved population of fainter CVs may exist in the core of the cluster, but only the brightest have been detected in our observation. In particular, we note that the source C1 has a luminosity (${L_{\mathrm{0.5-10~keV}}=9.5\times10^{32}\mathrm{~erg~s^{-1}}}$ if it belongs to the cluster) higher than the known luminosities of CVs \citep{Verbunt+94}, and could be composed of several sources.
After removing the contribution of C2 from the core global spectrum we found a spectral photon-index of $\Gamma=1.44\pm0.09$, consistent with CV emission \mbox{\citep[e.g.][]{BWO05}}, and only the brightest MSPs emission \citep{BGvdB05}. 
This supports the idea that the core emission is due to a majority of CVs.

In \object{NGC~4372} which is a less massive cluster, the estimation leads to less than one bright CV. If CVs are present in the cluster, they must be fainter than our limiting luminosity.

\paragraph{NGC~2808 -- C1.}
Source C1 is located in the core radius of \object{NGC~2808}. This source is hard, and is well fitted with a power law model or a bremsstrahlung model (see Table~\ref{table:2808fit}). These values are consistent with CV X-ray emission \mbox{\citep[e.g.][]{Richman96,BWO05}}.
C1 also has a UV counterpart (UV~222) proposed to be a CV by \citet{Dieball+05}, strengthening this hypothesis.
The spectrum is plotted in Fig.~\ref{fig:2808_C1_sp}. At 2.0~keV, a feature that at first sight could be an emission line is not well fitted with a Gaussian. It is possibly due to the fact that CV1 emission may be complex emission from several sources, as its luminosity is quite high for a single CV.

\paragraph{NGC~2808 -- C3, C4, C5.}
These sources are located in the half-mass radius and the colours indicate that they are likely to be CVs (Fig.~\ref{fig:2808hr} and \ref{fig:2808hrf}, and Table~\ref{table:2808src}). Their luminosity, if they are members of the cluster is consistent with this hypothesis.

\subsection{Millisecond pulsars in NGC~2808?}

Some GCs are known to harbour a large population of MSPs. In 47~Tuc, which is similar in mass to \object{NGC~2808}, a population of $\sim25$~MSPs is estimated \citep{Heinke+05}. 
The brightest MSPs have luminosities around our flux detection limit for \object{NGC~2808}.
However, we note that MSPs, which are hard X-ray sources, are difficult to distinguish from CVs at this luminosity \citep{BGvdB05}.
Therefore, our observations are not constraining for the presence of MSPs in \object{NGC~2808}.
For \object{NGC~4372}, their flux is too faint to be detected.

\subsection{Other sources possibly linked to NGC~2808}

We detect 5~sources with a hardness ratio below $-0.7$ in the flux-colour diagram presented in Fig.~\ref{fig:2808hrf}, and located ${\sim5-6\times r_{h[NGC2808]}}$ away.
Such sources could explain the possible excess of sources found in Sect.~\ref{section:members} if some are linked to the cluster.
There is a high probability that these sources are background sources, but if they are linked to the cluster, they could have been formed from primordial binaries and remain outside of the half-mass radius \citep{HAS07}.

\citet{DLSF93} presented a study of 44~RS~CVn systems (ABs) observed with the ROSAT observatory, and showed that their spectra are very soft with most of the emission below 2~keV. From their spectra (Table~\ref{table:2808fit}), colours and luminosity, sources~24 and~41 have similarities with the brightest ABs.

\citet{EH07} used a black body model to fit the soft excess below 2~keV found in the spectra of some intermediate polars, the hard emission of intermediate polars being fitted with a mekal model. Sources~22,~25 and~50 have soft spectra with a hard tail (Table~\ref{table:2808fit}), and could therefore be intermediate polars.

Some of the soft sources could also be active stars in the foreground. The X-ray flux here is consistent with an estimated distance of $\sim100$~pc.

\subsection{Looking for an IMBH in NGC~2808}

Following \citet{Trenti06}, \object{NGC~2808} is a good candidate for hosting an IMBH as the ratio $r_c/r_h$ is larger than the critical threshold of 0.3. This cluster may however not be sufficiently relaxed as the age is only 7.4 times the half-mass relaxation time \citep{Hurley07}.

If such an IMBH exists in \object{NGC~2808}, it should be located at the center of mass of the cluster due to mass segregation. We found no evidence for an X-ray source at this position but it may be below our detection limit.
We assume that the BH is fed by intracluster gas with a density of $\sim0.5\mathrm{~cm^{-3}}$ derived from \citet{PR01} with \object{NGC~2808} parameters. 
It should be noted, however, that a possible detection of $200\mathrm{~M_{\sun}}$ of neutral hydrogen in the
core of \object{NGC~2808} has been reported in the literature \citep{Faulkner+91}. If this is confirmed, it implies that the gas density is underestimated here.
In the same way as \citet{HTO03}, if we assume a BH of $1\,000\mathrm{~M_{\sun}}$ accreting at the full Bondi rate \citep{Bondi52} and an optically thick, geometrically thin disk \citep{SS73}, we found an X-ray accretion luminosity five orders of magnitude above our limiting luminosity.
However, the BH may be radiatively inefficient as for optically thin advection-dominated accretion flow \citep[ADAFs, see e.g.][]{NMQ98}. Following \citet{GHEM01}, our limiting luminosity implies an upper limit of $\sim290\mathrm{~M_{\sun}}$ for a central IMBH in \object{NGC~2808}.

\subsection{Other NGC~2808 sources}
\label{section:other2808}

\begin{table*}
\caption{\label{table:2808sum}      
\object{NGC~2808} core sources summary table. We give for each source the J2000 position with errors, the offset from the GC center, the luminosity in band 0.5--10~keV [$\times10^{-14}\mathrm{~erg~s^{-1}}$] and the hardness ratios ($HR$). The probable nature of the object is indicated as discussed in Sect.~\ref{section:discuss}. The $HR$ for two energy band $B1$ and $B2$ are defined by $HR=(B2-B1)/(B1+B2)$, $HR1$ for 0.5--1.5 and 1.5--3~keV energy bands, $HR2$ for 1.5--3 and 3--10~keV energy bands.}
\centering                          
\begin{tabular}{cc@{$^h$}c@{$^m$}c@{$^s$~~~}c@{\degr}c@{\arcmin}c@{\arcsec~~~}cccccc}
\hline\hline                 
 ID & \multicolumn{3}{c}{RA$_{2000}$} & \multicolumn{3}{c}{Dec$_{2000}$} & Error & Offset & Lum.  & $HR1$ & $HR2$ & Probable nature  \\
\hline                        
   C1 &  9&12&04.33 &  $-$64&51&49.09 &  3.47\arcsec & 0.19\arcmin &  9.5$\pm$1.7 &  $-$0.29 &   0.07 & CV    \\
   C2 &  9&12&01.65 &  $-$64&51&51.58 &  3.07\arcsec & 0.13\arcmin &  2.6$\pm$0.4 &  $-$0.96 &  $-$1.00 & qLMXB \\
   C3 &  9&12&00.58 &  $-$64&52&00.14 &  3.43\arcsec & 0.31\arcmin &  3.3$\pm$2.0 &  $-$0.18 &   0.20 & CV    \\
   C4 &  9&12&03.85 &  $-$64&52&04.50 &  4.47\arcsec & 0.32\arcmin &  3.2$\pm$2.1 &  $-$0.23 &  $-$0.20 & CV    \\
   C5 &  9&11&59.27 &  $-$64&51&46.74 &  4.47\arcsec & 0.35\arcmin &  3.0$\pm$1.8 &   0.33 &   0.08 & CV    \\
\hline                        
\end{tabular}
\end{table*}

\paragraph{NGC~2808 -- 1.}
This source is the brightest source in the field of view, located 8.5\arcmin\ away from the center of \object{NGC~2808}. A source was previously listed in the ASCA source catalogue \citep{Ueda+01} with a compatible position. The unabsorbed flux is lower but consistent with our detection (${F\mathrm{^{ASCA}_{0.7-7~keV}}=3.0\pm0.7\times10^{-13}\mathrm{~erg~cm^{-2}~s^{-1}}}$ and ${F\mathrm{^{XMM}_{0.7-7~keV}}=4.0\pm0.5\times10^{-13}\mathrm{~erg~cm^{-2}~s^{-1}}}$).

This source is well fitted with the absorbed power law model (APL+PO) used by \citet{Mainieri+07} for Seyfert~2 active galactic nuclei (AGN) spectra. It is composed of the Galactic absorption, an absorbed power law, and an unabsorbed power law to model the soft excess below 0.9~keV.
The lightcurve plotted in Fig.~\ref{fig:2808_1_lc} is typical of an AGN, showing short-term low amplitude variability \citep[see for instance ][]{Gliozzi+04}.
We note that a UV counterpart is found in the OM data.
These features are all consistent with an AGN, but an optical or infrared counterpart is needed to confirm the nature of this source.

\paragraph{NGC~2808 -- 5 and 13.}
These sources show very absorbed spectra, well fitted with the APL+PO model \citep{Mainieri+07}. These spectra could indicate Seyfert~2 AGN, as for source~1.

\paragraph{NGC~2808 -- 17.}
This source is located 3.2\arcmin\ away from the center of \object{NGC~2808}. We found a possible optical counterpart, \object{HD~79548} at RA$_{2000}$~$9^h11^m33.293^s$, Dec$_{2000}$~$-$64\degr51\arcmin03.28\arcsec. This star is an A0V star with magnitudes ${B=10.42}$ and ${V=10.15}$.
A UV counterpart is observed in the OM data with a magnitude of 10.95, consistent with the emission of \object{HD~79548}.
Using the distance and the visual magnitude of Vega, a well known A0V star (7.76~pc, ${V=0.03}$), we estimated a distance of $\sim800$~pc for \object{HD~79548}, and derived an X-ray luminosity of ${L_{\mathrm{0.5-10~keV}}=4.9\pm0.9\times10^{29}\mathrm{~erg~s^{-1}}}$.

The relation between the X-ray source and \object{HD~79548} is unclear and it is possible that we observe a background source aligned with \object{HD~79548}.
However, young A0V stars with an age less than $10^7$~yr can be such a bright X-ray source \mbox{\citep[e.g.][]{PDK06,TP95}}.
An active star as a companion is also consistent with the X-ray emission obtained \citep{BP03,Golub+83}. 
Optical spectroscopic observations of this source could help to determine the nature of this object.

\subsection{Peculiar NGC~4372 sources}

As the sources are distant from the core of the GC, they are more likely to be background sources.

\paragraph{NGC~4372 -- 1 and 2.}
These two soft sources appear to have similar parameters (Table~\ref{table:4372fit}). They are around 12\arcmin\ away from the center of \object{NGC~4372}, and are 2\arcmin\ away from each other.
Their spectra show a soft excess between 0.5 and 0.9~keV, possibly due to reprocessed emission of an absorbed AGN.

\paragraph{NGC~4372 -- 4.}
This source appears to be harder than the source R13 detected previously with ROSAT at the same location (within the error circles). With ROSAT, the photon-index corresponding to the hardness ratio is $\sim2$, and for our observation, it goes down to $\sim0.5$, if we assume the same absorption. If we assume the same photon-index for the two observations, then the absorption has doubled.

\paragraph{NGC~4372 -- 7, 8 and R10.}
In the contour map published by \citet{JVH96}, we can see some unresolved features at the positions of sources~7 and~8. We can estimate that their fluxes (Table~\ref{table:4372src}) have increased by a factor~12 and~4 for sources~7 and~8 respectively.
ROSAT source~R10 is not detected in our data. As we should have detected all ROSAT sources, it may have varied between the two observations by a factor 2 to become fainter than our limiting flux.


\section{Conclusions}

We have presented XMM-\textit{Newton} observations of the globular clusters \object{NGC~2808} and \object{NGC~4372}.

For \object{NGC~2808}, we have shown that the five central sources are likely to be linked to the cluster. One of these is very likely to be a qLMXB, and the emission of the remaining central sources is consistent with $20\pm10$ bright CVs (at luminosities above ${L_{\mathrm{0.5-10~keV}}=4.25\times10^{31}\mathrm{~erg~s^{-1}}}$), of which 4 are detected in our data. A summary of these sources is presented in Table~\ref{table:2808sum}.
We expect to resolve more objects in the core of \object{NGC~2808} in our Chandra observation \citep{Servillat+08}. 

For \object{NGC~4372}, we detect no sources in the half-mass radius, but the limiting luminosity of our observations is not constraining, in particular for a possible population of faint CVs.
We compared our sources outside the half-mass radius to previously detected sources with ROSAT in the field of view, and found a very good correlation for most of the sources.


\begin{acknowledgements}
This work is based on observations obtained with XMM-\textit{Newton}, an ESA science mission with instruments and contributions directly funded by ESA Member States and NASA.
We thank the CNES for support of the operational phase of the mission.
This research has made use of the SIMBAD database, operated at the CDS, Strasbourg, France.
We are grateful to J. Grindlay whose comments have helped to improve this paper.
\end{acknowledgements}



\begin{thebibliography}{71}
\expandafter\ifx\csname natexlab\endcsname\relax\def\natexlab#1{#1}\fi

\bibitem[{{Alcaino} {et~al.}(1991){Alcaino}, {Liller}, {Alvarado}, \&
  {Wenderoth}}]{Alcaino+91}
{Alcaino}, G., {Liller}, W., {Alvarado}, F., \& {Wenderoth}, E. 1991, \aj, 102,
  159

\bibitem[{{Arnaud}(1996)}]{Arnaud96}
{Arnaud}, K.~A. 1996, in ASP Conf. Ser. 101: Astronomical Data Analysis
  Software and Systems V, ed. G.~H. {Jacoby} \& J.~{Barnes}, 17

\bibitem[{{Baskill} {et~al.}(2005){Baskill}, {Wheatley}, \& {Osborne}}]{BWO05}
{Baskill}, D.~S., {Wheatley}, P.~J., \& {Osborne}, J.~P. 2005, \mnras, 357, 626

\bibitem[{{Bedin} {et~al.}(2000){Bedin}, {Piotto}, {Zoccali}, {Stetson},
  {Saviane}, {Cassisi}, \& {Bono}}]{Bedin+00}
{Bedin}, L.~R., {Piotto}, G., {Zoccali}, M., {et~al.} 2000, \aap, 363, 159

\bibitem[{{Bogdanov} {et~al.}(2005){Bogdanov}, {Grindlay}, \& {van den
  Berg}}]{BGvdB05}
{Bogdanov}, S., {Grindlay}, J.~E., \& {van den Berg}, M. 2005, \apj, 630, 1029

\bibitem[{{Bondi}(1952)}]{Bondi52}
{Bondi}, H. 1952, \mnras, 112, 195

\bibitem[{{Briggs} \& {Pye}(2003)}]{BP03}
{Briggs}, K.~R. \& {Pye}, J.~P. 2003, \mnras, 345, 714

\bibitem[{{Burstein} \& {Heiles}(1978)}]{BH78}
{Burstein}, D. \& {Heiles}, C. 1978, \aplett, 19, 69

\bibitem[{{Carretta} {et~al.}(2006){Carretta}, {Bragaglia}, {Gratton}, {Leone},
  {Recio-Blanco}, \& {Lucatello}}]{Carretta+06}
{Carretta}, E., {Bragaglia}, A., {Gratton}, R.~G., {et~al.} 2006, \aap, 450,
  523

\bibitem[{{Carretta} {et~al.}(2000){Carretta}, {Gratton}, {Clementini}, \&
  {Fusi Pecci}}]{CGCFP00}
{Carretta}, E., {Gratton}, R.~G., {Clementini}, G., \& {Fusi Pecci}, F. 2000,
  \apj, 533, 215

\bibitem[{{Cash}(1979)}]{Cash79}
{Cash}, W. 1979, \apj, 228, 939

\bibitem[{{Colpi} {et~al.}(2003){Colpi}, {Mapelli}, \& {Possenti}}]{CMP03}
{Colpi}, M., {Mapelli}, M., \& {Possenti}, A. 2003, \apj, 599, 1260

\bibitem[{{D'Amico} {et~al.}(2002){D'Amico}, {Possenti}, {Fici}, {Manchester},
  {Lyne}, {Camilo}, \& {Sarkissian}}]{DAmico+02}
{D'Amico}, N., {Possenti}, A., {Fici}, L., {et~al.} 2002, \apjl, 570, L89

\bibitem[{{Dempsey} {et~al.}(1993){Dempsey}, {Linsky}, {Schmitt}, \&
  {Fleming}}]{DLSF93}
{Dempsey}, R.~C., {Linsky}, J.~L., {Schmitt}, J.~H.~M.~M., \& {Fleming}, T.~A.
  1993, \apj, 413, 333

\bibitem[{{Dieball} {et~al.}(2005){Dieball}, {Knigge}, {Zurek}, {Shara}, \&
  {Long}}]{Dieball+05}
{Dieball}, A., {Knigge}, C., {Zurek}, D.~R., {Shara}, M.~M., \& {Long}, K.~S.
  2005, \apj, 625, 156

\bibitem[{{Edmonds} {et~al.}(2003){Edmonds}, {Gilliland}, {Heinke}, \&
  {Grindlay}}]{EGHG03}
{Edmonds}, P.~D., {Gilliland}, R.~L., {Heinke}, C.~O., \& {Grindlay}, J.~E.
  2003, \apj, 596, 1177

\bibitem[{{Ehle} {et~al.}(2006){Ehle}, {Breitfellner}, {Gonzalez Riestra},
  {Guainazzi}, {Loiseau}, {Rodriguez}, {Santos-Lleo}, {Schartel}, {Tomas},
  {Verdugo}, \& {Dahlem}}]{XMM-UHB}
{Ehle}, M., {Breitfellner}, M., {Gonzalez Riestra}, R., {et~al.} 2006,
  XMM-Newton Users' Handbook v2.4

\bibitem[{{Evans} \& {Hellier}(2007)}]{EH07}
{Evans}, P.~A. \& {Hellier}, C. 2007, \apj, 663, 1277

\bibitem[{{Faulkner} {et~al.}(1991){Faulkner}, {Scott}, {Wood}, \&
  {Wright}}]{Faulkner+91}
{Faulkner}, D.~J., {Scott}, T.~R., {Wood}, P.~R., \& {Wright}, A.~E. 1991,
  \apjl, 374, L45

\bibitem[{{Gebhardt} {et~al.}(2002){Gebhardt}, {Rich}, \& {Ho}}]{GRH02}
{Gebhardt}, K., {Rich}, R.~M., \& {Ho}, L.~C. 2002, \apjl, 578, L41

\bibitem[{{Gebhardt} {et~al.}(2005){Gebhardt}, {Rich}, \& {Ho}}]{GRH05}
{Gebhardt}, K., {Rich}, R.~M., \& {Ho}, L.~C. 2005, \apj, 634, 1093

\bibitem[{{Gendre} {et~al.}(2003{\natexlab{a}}){Gendre}, {Barret}, \&
  {Webb}}]{GBW03}
{Gendre}, B., {Barret}, D., \& {Webb}, N.~A. 2003{\natexlab{a}}, \aap, 403, L11

\bibitem[{{Gendre} {et~al.}(2003{\natexlab{b}}){Gendre}, {Barret}, \&
  {Webb}}]{GBW03b}
{Gendre}, B., {Barret}, D., \& {Webb}, N.~A. 2003{\natexlab{b}}, \aap, 400, 521

\bibitem[{{Gerssen} {et~al.}(2002){Gerssen}, {van der Marel}, {Gebhardt},
  {Guhathakurta}, {Peterson}, \& {Pryor}}]{Gerssen+02}
{Gerssen}, J., {van der Marel}, R.~P., {Gebhardt}, K., {et~al.} 2002, \aj, 124,
  3270

\bibitem[{{Gliozzi} {et~al.}(2004){Gliozzi}, {Sambruna}, {Brandt}, {Mushotzky},
  \& {Eracleous}}]{Gliozzi+04}
{Gliozzi}, M., {Sambruna}, R.~M., {Brandt}, W.~N., {Mushotzky}, R., \&
  {Eracleous}, M. 2004, \aap, 413, 139

\bibitem[{{Golub} {et~al.}(1983){Golub}, {Harnden}, {Maxson}, {Vaiana}, {Snow},
  {Rosner}, \& {Cash}}]{Golub+83}
{Golub}, L., {Harnden}, Jr., F.~R., {Maxson}, C.~W., {et~al.} 1983, \apj, 271,
  264

\bibitem[{{Grindlay} {et~al.}(2001){Grindlay}, {Heinke}, {Edmonds}, \&
  {Murray}}]{GHEM01}
{Grindlay}, J.~E., {Heinke}, C., {Edmonds}, P.~D., \& {Murray}, S.~S. 2001,
  Science, 292, 2290

\bibitem[{{Harris}(1974)}]{Harris74}
{Harris}, W.~E. 1974, \apjl, 192, L161

\bibitem[{{Harris}(1996)}]{Harris96}
{Harris}, W.~E. 1996, \aj, 112, 1487

\bibitem[{{Hasinger} {et~al.}(2005){Hasinger}, {Miyaji}, \& {Schmidt}}]{HMS05}
{Hasinger}, G., {Miyaji}, T., \& {Schmidt}, M. 2005, \aap, 441, 417

\bibitem[{{Heinke} {et~al.}(2005){Heinke}, {Grindlay}, {Edmonds}, {Cohn},
  {Lugger}, {Camilo}, {Bogdanov}, \& {Freire}}]{Heinke+05}
{Heinke}, C.~O., {Grindlay}, J.~E., {Edmonds}, P.~D., {et~al.} 2005, \apj, 625,
  796

\bibitem[{{Heinke} {et~al.}(2003{\natexlab{a}}){Heinke}, {Grindlay}, {Lloyd},
  \& {Edmonds}}]{HGLE03}
{Heinke}, C.~O., {Grindlay}, J.~E., {Lloyd}, D.~A., \& {Edmonds}, P.~D.
  2003{\natexlab{a}}, \apj, 588, 452

\bibitem[{{Heinke} {et~al.}(2003{\natexlab{b}}){Heinke}, {Grindlay}, {Lugger},
  {Cohn}, {Edmonds}, {Lloyd}, \& {Cool}}]{Heinke+03}
{Heinke}, C.~O., {Grindlay}, J.~E., {Lugger}, P.~M., {et~al.}
  2003{\natexlab{b}}, \apj, 598, 501

\bibitem[{{Heinke} {et~al.}(2006){Heinke}, {Wijnands}, {Cohn}, {Lugger},
  {Grindlay}, {Pooley}, \& {Lewin}}]{Heinke+06}
{Heinke}, C.~O., {Wijnands}, R., {Cohn}, H.~N., {et~al.} 2006, \apj, 651, 1098

\bibitem[{{Ho} {et~al.}(2003){Ho}, {Terashima}, \& {Okajima}}]{HTO03}
{Ho}, L.~C., {Terashima}, Y., \& {Okajima}, T. 2003, \apjl, 587, L35

\bibitem[{{Hurley}(2007)}]{Hurley07}
{Hurley}, J.~R. 2007, \mnras, 379, 93

\bibitem[{{Hurley} {et~al.}(2007){Hurley}, {Aarseth}, \& {Shara}}]{HAS07}
{Hurley}, J.~R., {Aarseth}, S.~J., \& {Shara}, M.~M. 2007, \apj, 665, 707

\bibitem[{{Hut} {et~al.}(1992){Hut}, {McMillan}, {Goodman}, {Mateo}, {Phinney},
  {Pryor}, {Richer}, {Verbunt}, \& {Weinberg}}]{Hut+92}
{Hut}, P., {McMillan}, S., {Goodman}, J., {et~al.} 1992, \pasp, 104, 981

\bibitem[{{Hut} {et~al.}(2003){Hut}, {Shara}, {Aarseth}, {Klessen}, {Lombardi},
  {Makino}, {McMillan}, {Pols}, {Teuben}, \& {Webbink}}]{Hut+03}
{Hut}, P., {Shara}, M.~M., {Aarseth}, S.~J., {et~al.} 2003, New Astronomy, 8,
  337

\bibitem[{{Ivanova} {et~al.}(2006){Ivanova}, {Heinke}, {Rasio}, {Taam},
  {Belczynski}, \& {Fregeau}}]{Ivanova+06}
{Ivanova}, N., {Heinke}, C.~O., {Rasio}, F.~A., {et~al.} 2006, \mnras, 372,
  1043

\bibitem[{{Johnston} {et~al.}(1996){Johnston}, {Verbunt}, \&
  {Hasinger}}]{JVH96}
{Johnston}, H.~M., {Verbunt}, F., \& {Hasinger}, G. 1996, \aap, 309, 116

\bibitem[{{Kaluzny} \& {Krzeminski}(1993)}]{KK93}
{Kaluzny}, J. \& {Krzeminski}, W. 1993, \mnras, 264, 785

\bibitem[{{Lewin} \& {Joss}(1983)}]{1983adsx.conf...41L}
{Lewin}, W.~H.~G. \& {Joss}, P.~C. 1983, in Accretion-Driven Stellar X-ray
  Sources, ed. W.~H.~G. {Lewin} \& E.~P.~J. {van den Heuvel}, 41

\bibitem[{{Lightman} \& {Grindlay}(1982)}]{LG82}
{Lightman}, A.~P. \& {Grindlay}, J.~E. 1982, \apj, 262, 145

\bibitem[{{Lockman} {et~al.}(1986){Lockman}, {Jahoda}, \& {McCammon}}]{LJM86}
{Lockman}, F.~J., {Jahoda}, K., \& {McCammon}, D. 1986, \apj, 302, 432

\bibitem[{{Loiseau}(2006)}]{SAS-USG}
{Loiseau}, N. 2006, User's Guide to the XMM-Newton Science Analysis System v4.0

\bibitem[{{Maccarone} {et~al.}(2007){Maccarone}, {Kundu}, {Zepf}, \&
  {Rhode}}]{Maccarone+07}
{Maccarone}, T.~J., {Kundu}, A., {Zepf}, S.~E., \& {Rhode}, K.~L. 2007, \nat,
  445, 183

\bibitem[{{Mainieri} {et~al.}(2007){Mainieri}, {Hasinger}, {Cappelluti},
  {Brusa}, {Brunner}, {Civano}, {Comastri}, {Elvis}, {Finoguenov}, {Fiore},
  {Gilli}, {Lehmann}, {Silverman}, {Tasca}, {Vignali}, {Zamorani},
  {Schinnerer}, {Impey}, {Trump}, {Lilly}, {Maier}, {Griffiths}, {Miyaji},
  {Capak}, {Koekemoer}, {Scoville}, {Shopbell}, \& {Taniguchi}}]{Mainieri+07}
{Mainieri}, V., {Hasinger}, G., {Cappelluti}, N., {et~al.} 2007, \apjs, 172,
  368

\bibitem[{{Mukai}(1993)}]{Mukai93}
{Mukai}, K. 1993, Legacy, vol.~3, p.21-31, 3, 21

\bibitem[{{Narayan} {et~al.}(1998){Narayan}, {Mahadevan}, \&
  {Quataert}}]{NMQ98}
{Narayan}, R., {Mahadevan}, R., \& {Quataert}, E. 1998, in Theory of Black Hole
  Accretion Disks, ed. M.~A. {Abramowicz}, G.~{Bjornsson}, \& J.~E. {Pringle},
  148

\bibitem[{{Pease} {et~al.}(2006){Pease}, {Drake}, \& {Kashyap}}]{PDK06}
{Pease}, D.~O., {Drake}, J.~J., \& {Kashyap}, V.~L. 2006, \apj, 636, 426

\bibitem[{{Pfahl} \& {Rappaport}(2001)}]{PR01}
{Pfahl}, E. \& {Rappaport}, S. 2001, \apj, 550, 172

\bibitem[{{Pietrukowicz} {et~al.}(2005){Pietrukowicz}, {Kaluzny}, {Thompson},
  {Jaroszynski}, {Schwarzenberg-Czerny}, {Krzeminski}, \&
  {Pych}}]{Pietrukowicz+05}
{Pietrukowicz}, P., {Kaluzny}, J., {Thompson}, I.~B., {et~al.} 2005, Acta
  Astronomica, 55, 261

\bibitem[{{Piotto} {et~al.}(2007){Piotto}, {Bedin}, {Anderson}, {King},
  {Cassisi}, {Milone}, {Villanova}, {Pietrinferni}, \& {Renzini}}]{Piotto+07}
{Piotto}, G., {Bedin}, L.~R., {Anderson}, J., {et~al.} 2007, \apjl, 661, L53

\bibitem[{{Pooley} \& {Hut}(2006)}]{PH06}
{Pooley}, D. \& {Hut}, P. 2006, \apjl, 646, L143

\bibitem[{{Pooley} {et~al.}(2003){Pooley}, {Lewin}, {Anderson}, {Baumgardt},
  {Filippenko}, {Gaensler}, {Homer}, {Hut}, {Kaspi}, {Makino}, {Margon},
  {McMillan}, {Portegies Zwart}, {van der Klis}, \& {Verbunt}}]{Pooley+03}
{Pooley}, D., {Lewin}, W.~H.~G., {Anderson}, S.~F., {et~al.} 2003, \apjl, 591,
  L131

\bibitem[{{Richman}(1996)}]{Richman96}
{Richman}, H.~R. 1996, \apj, 462, 404

\bibitem[{{Servillat} {et~al.}(2008){Servillat}, {Webb}, {Barret},
  {Cornelisse}, {Dieball}, {Knigge}, {Long}, {Shara}, \&
  {Zurek}}]{Servillat+08}
{Servillat}, M., {Webb}, N.~A., {Barret}, D., {et~al.} 2008, in IAU Symposium
  246, Dynamical Evolution of Dense Stellar Systems, ed. E.~{Vesperini},
  M.~{Giersz}, \& A.~{Sills}, submitted

\bibitem[{{Shakura} \& {Syunyaev}(1973)}]{SS73}
{Shakura}, N.~I. \& {Syunyaev}, R.~A. 1973, \aap, 24, 337

\bibitem[{{Tout} \& {Pringle}(1995)}]{TP95}
{Tout}, C.~A. \& {Pringle}, J.~E. 1995, \mnras, 272, 528

\bibitem[{{Trenti}(2006)}]{Trenti06}
{Trenti}, M. 2006, \mnras, submitted (astro-ph/0612040)

\bibitem[{{Trenti} {et~al.}(2007){Trenti}, {Ardi}, {Mineshige}, \&
  {Hut}}]{Trenti+07}
{Trenti}, M., {Ardi}, E., {Mineshige}, S., \& {Hut}, P. 2007, \mnras, 374, 857

\bibitem[{{Ueda} {et~al.}(2001){Ueda}, {Ishisaki}, {Takahashi}, {Makishima}, \&
  {Ohashi}}]{Ueda+01}
{Ueda}, Y., {Ishisaki}, Y., {Takahashi}, T., {Makishima}, K., \& {Ohashi}, T.
  2001, \apjs, 133, 1

\bibitem[{{Verbunt} \& {Hut}(1987)}]{VH87}
{Verbunt}, F. \& {Hut}, P. 1987, in IAU Symp. 125: The Origin and Evolution of
  Neutron Stars, ed. D.~J. {Helfand} \& J.-H. {Huang}, 187

\bibitem[{{Verbunt} {et~al.}(1994){Verbunt}, {Johnston}, {Hasinger}, {Belloni},
  \& {Bunk}}]{Verbunt+94}
{Verbunt}, F., {Johnston}, H., {Hasinger}, G., {Belloni}, T., \& {Bunk}, W.
  1994, in Astronomical Society of the Pacific Conference Series, Vol.~56,
  Interacting Binary Stars, ed. A.~W. {Shafter}, 244

\bibitem[{{Voges} {et~al.}(1999){Voges}, {Aschenbach}, {Boller},
  {Br{\"a}uninger}, {Briel}, {Burkert}, {Dennerl}, {Englhauser}, {Gruber},
  {Haberl}, {Hartner}, {Hasinger}, {K{\"u}rster}, {Pfeffermann}, {Pietsch},
  {Predehl}, {Rosso}, {Schmitt}, {Tr{\"u}mper}, \& {Zimmermann}}]{Voges+99}
{Voges}, W., {Aschenbach}, B., {Boller}, T., {et~al.} 1999, \aap, 349, 389

\bibitem[{Webb \& Barret(2007)}]{WB07}
Webb, N.~A. \& Barret, D. 2007, \apj

\bibitem[{{Webb} {et~al.}(2004){Webb}, {Serre}, {Gendre}, {Barret}, {Lasota},
  \& {Rizzi}}]{Webb+04}
{Webb}, N.~A., {Serre}, D., {Gendre}, B., {et~al.} 2004, \aap, 424, 133

\bibitem[{{Webb} {et~al.}(2006){Webb}, {Wheatley}, \& {Barret}}]{WWB06}
{Webb}, N.~A., {Wheatley}, P.~J., \& {Barret}, D. 2006, \aap, 445, 155

\bibitem[{{Yang} {et~al.}(2003){Yang}, {Mushotzky}, {Barger}, {Cowie},
  {Sanders}, \& {Steffen}}]{Yang+03}
{Yang}, Y., {Mushotzky}, R.~F., {Barger}, A.~J., {et~al.} 2003, \apjl, 585, L85

\bibitem[{{Zavlin} {et~al.}(1996){Zavlin}, {Pavlov}, \& {Shibanov}}]{ZPS96}
{Zavlin}, V.~E., {Pavlov}, G.~G., \& {Shibanov}, Y.~A. 1996, \aap, 315, 141

\end{thebibliography}


\onllongtab{2}{
\begin{longtable}{cc@{$^h$}c@{$^m$}c@{$^s$~~~~}c@{\degr}c@{\arcmin}c@{\arcsec~~~~}c@{\arcsec~~~~}cccccc}
\caption{\label{table:2808src}
\object{NGC~2808} X-ray source properties. The columns contain the ID of the source, the position (RA$_{2000}$ and Dec$_{2000}$) with the error, the rate [$\times$~10$^{-3}$~count~s$^{-1}$], the flux converted with ECFs in band 0.5--10~keV [$\times$~10$^{-14}$~erg~cm$^{-2}$~s$^{-1}$] and the hardness ratios ($HR$). The $HR$ for two energy band $B1$ and $B2$ are defined by $HR=(B2-B1)/(B1+B2)$, $HR1$ for 0.5--1.5 and 1.5--3~keV energy bands, $HR2$ for 1.5--3 and 3--10~keV energy bands.} \\
\hline\hline
 ID & \multicolumn{3}{c}{RA$_{2000}$} & \multicolumn{3}{c}{Dec$_{2000}$} & \multicolumn{1}{c}{Error} & Rate & Flux & $HR1$ & $HR2$ \\
\hline
\endfirsthead
\caption{continued.}\\
\hline\hline
 ID & \multicolumn{3}{c}{RA$_{2000}$} & \multicolumn{3}{c}{Dec$_{2000}$} & \multicolumn{1}{l}{Error} & Rate & Flux & $HR1$ & $HR2$ \\
\hline
\endhead
\hline
\endfoot
   C1 &  9&12&04.33 &  $-$64&51&49.09 &  3.47 &  29.1~$\pm$~5.8 &  9.1~$\pm$~4.2 &  $-$0.29 &  0.07 \\
   C2 &  9&12&01.65 &  $-$64&51&51.58 &  3.07 &  18.5~$\pm$~4.4 &  2.5~$\pm$~1.5 &  $-$0.96 &  $-$1.00 \\
   C3 &  9&12&00.58 &  $-$64&52&00.14 &  3.43 &  10.4~$\pm$~2.4 &  3.0~$\pm$~1.8 &  $-$0.18 &  0.20 \\
   C4 &  9&12&03.85 &  $-$64&52&04.50 &  4.47 &  7.9~$\pm$~2.3 &  2.9~$\pm$~1.9 &  $-$0.23 &  $-$0.20 \\
   C5 &  9&11&59.27 &  $-$64&51&46.74 &  4.47 &  5.9~$\pm$~1.8 &  2.7~$\pm$~1.6 &  0.33 &  0.08 \\
\hline
    1 &  9&10&49.68 &  $-$64&48&15.73 &  2.73 &  223.9~$\pm$~5.1 &  71.4~$\pm$~4.2 &  $-$0.07 &  $-$0.04 \\
    2 &  9&11&28.05 &  $-$64&50&37.34 &  2.73 &  72.8~$\pm$~2.3 &  17.0~$\pm$~1.5 &  $-$0.28 &  $-$0.26 \\
    3 &  9&11&18.94 &  $-$64&51&00.39 &  2.73 &  58.2~$\pm$~2.1 &  9.7~$\pm$~1.0 &  $-$0.54 &  $-$0.36 \\
    5 &  9&12&56.89 &  $-$64&52&28.86 &  2.75 &  48.7~$\pm$~2.9 &  30.1~$\pm$~4.3 &  0.78 &  0.46 \\
    6 &  9&13&01.41 &  $-$64&52&17.62 &  2.78 &  17.9~$\pm$~1.2 &  3.7~$\pm$~0.8 &  $-$0.27 &  $-$0.53 \\
    7 &  9&12&18.36 &  $-$64&48&41.90 &  2.77 &  15.7~$\pm$~1.1 &  3.1~$\pm$~0.6 &  $-$0.41 &  $-$0.37 \\
    8 &  9&13&41.37 &  $-$64&44&30.49 &  2.79 &  38.7~$\pm$~2.9 &  9.4~$\pm$~2.2 &  $-$0.31 &  $-$0.24 \\
    9 &  9&12&49.16 &  $-$64&44&49.84 &  2.80 &  20.9~$\pm$~1.6 &  6.1~$\pm$~1.3 &  $-$0.33 &  0.04 \\
   10 &  9&13&27.75 &  $-$64&43&20.61 &  2.79 &  39.3~$\pm$~2.9 &  14.5~$\pm$~2.7 &  0.01 &  0.07 \\
   11 &  9&12&25.55 &  $-$65&01&03.40 &  2.81 &  20.9~$\pm$~1.7 &  2.0~$\pm$~0.5 &  $-$0.85 &  $-$0.77 \\
   12 &  9&11&37.75 &  $-$64&44&04.26 &  2.80 &  18.5~$\pm$~1.6 &  4.6~$\pm$~1.2 &  $-$0.27 &  $-$0.19 \\
   13 &  9&11&26.44 &  $-$64&53&15.84 &  2.80 &  10.7~$\pm$~0.9 &  5.4~$\pm$~1.0 &  0.48 &  0.13 \\
   14 &  9&12&49.98 &  $-$64&56&13.02 &  2.79 &  11.2~$\pm$~1.0 &  3.4~$\pm$~0.8 &  0.09 &  $-$0.23 \\
   15 &  9&12&51.95 &  $-$65&02&12.43 &  2.81 &  19.3~$\pm$~1.8 &  5.3~$\pm$~1.4 &  $-$0.16 &  $-$0.17 \\
   16 &  9&13&37.17 &  $-$64&52&01.33 &  2.83 &  17.0~$\pm$~1.6 &  6.6~$\pm$~1.5 &  0.08 &  0.12 \\
   17 &  9&11&33.49 &  $-$64&51&04.13 &  2.82 &  8.4~$\pm$~0.8 &  1.6~$\pm$~0.5 &  $-$0.51 &  $-$0.33 \\
   18 &  9&13&01.98 &  $-$64&44&56.08 &  2.90 &  13.6~$\pm$~1.6 &  4.9~$\pm$~1.4 &  $-$0.01 &  $-$0.07 \\
   19 &  9&10&57.78 &  $-$64&49&36.80 &  2.85 &  16.4~$\pm$~1.9 &  4.0~$\pm$~1.1 &  $-$0.35 &  $-$0.14 \\
   20 &  9&13&28.80 &  $-$64&59&18.90 &  2.85 &  12.4~$\pm$~1.5 &  1.9~$\pm$~0.9 &  $-$0.90 &  $-$0.07 \\
   21 &  9&12&42.59 &  $-$65&01&25.98 &  2.83 &  16.3~$\pm$~1.6 &  5.3~$\pm$~1.4 &  $-$0.26 &  0.13 \\
   22 &  9&12&35.07 &  $-$64&53&17.99 &  2.86 &  6.8~$\pm$~0.7 &  1.2~$\pm$~0.4 &  $-$0.65 &  $-$0.25 \\
   23 &  9&12&36.92 &  $-$64&45&45.28 &  2.90 &  9.8~$\pm$~1.1 &  3.1~$\pm$~0.9 &  $-$0.38 &  0.15 \\
   24 &  9&12&01.49 &  $-$64&56&14.25 &  2.85 &  5.8~$\pm$~0.7 &  0.9~$\pm$~0.4 &  $-$0.99 &  $-$0.38 \\
   25 &  9&11&28.92 &  $-$64&55&23.47 &  2.85 &  6.4~$\pm$~0.8 &  0.9~$\pm$~0.4 &  $-$0.67 &  $-$0.61 \\
   26 &  9&12&26.97 &  $-$64&48&49.42 &  2.88 &  6.6~$\pm$~0.8 &  1.3~$\pm$~0.5 &  $-$0.39 &  $-$0.28 \\
   27 &  9&13&23.49 &  $-$64&51&04.73 &  2.87 &  9.8~$\pm$~1.2 &  2.8~$\pm$~1.0 &  $-$0.60 &  0.34 \\
   28 &  9&11&57.53 &  $-$65&02&53.79 &  2.94 &  9.7~$\pm$~1.3 &  1.8~$\pm$~0.9 &  $-$0.52 &  $-$0.17 \\
   29 &  9&12&02.18 &  $-$64&55&08.08 &  2.87 &  5.8~$\pm$~0.7 &  1.6~$\pm$~0.5 &  $-$0.43 &  0.07 \\
   30 &  9&11&06.52 &  $-$65&01&56.63 &  2.89 &  10.0~$\pm$~1.4 &  3.0~$\pm$~1.2 &  0.17 &  $-$0.39 \\
   31 &  9&13&12.59 &  $-$64&52&57.75 &  2.91 &  5.9~$\pm$~0.8 &  1.1~$\pm$~0.6 &  $-$0.31 &  $-$0.68 \\
   32 &  9&09&55.01 &  $-$64&51&10.46 &  3.03 &  15.8~$\pm$~2.5 &  3.9~$\pm$~2.0 &  $-$0.40 &  $-$0.24 \\
   33 &  9&10&42.43 &  $-$64&58&00.14 &  2.98 &  10.4~$\pm$~1.6 &  2.9~$\pm$~1.2 &  $-$0.10 &  $-$0.30 \\
   34 &  9&11&37.72 &  $-$64&46&02.15 &  2.99 &  6.8~$\pm$~0.9 &  1.6~$\pm$~0.7 &  $-$0.15 &  $-$0.39 \\
   35 &  9&10&26.42 &  $-$64&55&29.70 &  2.94 &  10.1~$\pm$~1.4 &  4.0~$\pm$~1.5 &  $-$0.31 &  0.13 \\
   36 &  9&12&42.88 &  $-$64&44&34.46 &  3.03 &  7.3~$\pm$~1.1 &  2.3~$\pm$~1.0 &  $-$0.11 &  0.08 \\
   37 &  9&12&33.26 &  $-$64&50&28.09 &  2.93 &  4.9~$\pm$~0.7 &  2.2~$\pm$~0.7 &  0.11 &  0.06 \\
   38 &  9&11&17.38 &  $-$64&46&54.15 &  2.90 &  9.0~$\pm$~1.4 &  2.3~$\pm$~1.0 &  $-$0.34 &  0.00 \\
   39 &  9&10&27.64 &  $-$65&00&59.56 &  3.03 &  11.6~$\pm$~1.9 &  3.6~$\pm$~1.9 &  $-$0.13 &  $-$0.20 \\
   40 &  9&12&54.61 &  $-$64&41&00.71 &  3.17 &  12.7~$\pm$~2.0 &  3.6~$\pm$~1.7 &  0.07 &  0.07 \\
   41 &  9&12&34.56 &  $-$64&53&47.62 &  2.98 &  3.3~$\pm$~0.5 &  0.4~$\pm$~0.2 &  $-$0.95 &  $-$0.34 \\
   42 &  9&13&49.91 &  $-$64&44&30.75 &  3.20 &  9.2~$\pm$~1.8 &  2.2~$\pm$~1.4 &  $-$0.90 &  0.63 \\
   43 &  9&12&09.20 &  $-$65&00&14.31 &  3.06 &  5.6~$\pm$~0.9 &  2.5~$\pm$~0.9 &  $-$0.28 &  0.17 \\
   44 &  9&12&10.38 &  $-$64&42&45.63 &  2.96 &  6.6~$\pm$~1.0 &  1.2~$\pm$~0.6 &  $-$0.61 &  $-$0.61 \\
   45 &  9&11&34.52 &  $-$64&55&58.33 &  2.95 &  4.6~$\pm$~0.7 &  0.8~$\pm$~0.4 &  $-$0.07 &  $-$0.72 \\
   46 &  9&12&26.58 &  $-$64&52&57.28 &  2.91 &  3.6~$\pm$~0.6 &  0.8~$\pm$~0.4 &  $-$0.41 &  $-$0.09 \\
   47 &  9&11&11.98 &  $-$64&55&04.46 &  2.95 &  5.4~$\pm$~0.8 &  1.1~$\pm$~0.6 &  0.11 &  $-$0.74 \\
   48 &  9&13&37.18 &  $-$64&59&23.29 &  3.01 &  9.7~$\pm$~1.6 &  3.6~$\pm$~1.5 &  $-$0.46 &  0.40 \\
   49 &  9&10&43.80 &  $-$64&57&21.68 &  3.10 &  6.4~$\pm$~1.1 &  0.9~$\pm$~0.6 &  $-$0.43 &  $-$0.73 \\
   50 &  9&12&39.39 &  $-$64&53&30.91 &  3.13 &  3.2~$\pm$~0.6 &  0.6~$\pm$~0.4 &  $-$0.74 &  $-$0.28 \\
   51 &  9&11&19.71 &  $-$65&01&39.27 &  3.18 &  6.2~$\pm$~1.2 &  2.0~$\pm$~1.2 &  0.11 &  $-$0.08 \\
   52 &  9&10&10.14 &  $-$64&49&32.70 &  3.36 &  9.9~$\pm$~1.9 &  3.4~$\pm$~1.8 &  $-$0.37 &  0.24 \\
   53 &  9&11&27.99 &  $-$64&45&16.78 &  3.10 &  5.1~$\pm$~1.1 &  2.5~$\pm$~1.1 &  $-$0.15 &  0.35 \\
   54 &  9&13&48.16 &  $-$64&49&13.83 &  3.33 &  5.8~$\pm$~1.5 &  1.7~$\pm$~1.1 &  0.09 &  $-$0.44 \\
   55 &  9&10&49.56 &  $-$64&53&04.38 &  3.01 &  4.6~$\pm$~0.8 &  1.4~$\pm$~0.8 &  $-$0.25 &  $-$0.34 \\
   56 &  9&12&44.02 &  $-$64&57&45.26 &  3.17 &  4.0~$\pm$~0.8 &  1.1~$\pm$~0.6 &  0.04 &  $-$0.36 \\
   57 &  9&11&31.38 &  $-$64&40&31.48 &  3.20 &  7.6~$\pm$~1.6 &  2.8~$\pm$~1.6 &  $-$0.10 &  $-$0.12 \\
   58 &  9&10&06.13 &  $-$64&42&24.11 &  3.12 &  11.0~$\pm$~2.6 &  6.0~$\pm$~2.4 &  $-$0.50 &  0.36 \\
   59 &  9&11&08.53 &  $-$64&51&13.89 &  3.13 &  3.3~$\pm$~0.7 &  1.4~$\pm$~0.6 &  0.47 &  $-$0.27 \\
   60 &  9&11&13.27 &  $-$64&44&01.37 &  3.22 &  6.1~$\pm$~1.2 &  2.0~$\pm$~1.1 &  $-$0.40 &  0.30 \\
   61 &  9&11&54.16 &  $-$65&02&07.50 &  3.09 &  4.6~$\pm$~1.0 &  1.9~$\pm$~0.9 &  0.46 &  $-$0.36 \\
   62 &  9&12&51.81 &  $-$64&52&10.04 &  3.24 &  2.6~$\pm$~0.6 &  0.6~$\pm$~0.5 &  0.30 &  $-$0.41 \\
   63 &  9&13&57.26 &  $-$64&49&07.89 &  3.30 &  3.2~$\pm$~0.8 &  1.7~$\pm$~1.0 &  $-$0.43 &  $-$0.50 \\
   64 &  9&13&07.10 &  $-$64&38&42.62 &  3.86 &  8.1~$\pm$~2.2 &  5.1~$\pm$~2.1 &  $-$0.12 &  0.24 \\
   65 &  9&14&07.11 &  $-$64&56&07.10 &  3.57 &  3.7~$\pm$~0.8 &  1.4~$\pm$~0.9 &  $-$0.77 &  $-$0.29 \\
   66 &  9&11&37.91 &  $-$64&48&56.44 &  3.16 &  2.8~$\pm$~0.6 &  0.7~$\pm$~0.4 &  $-$0.81 &  0.50 \\
   67 &  9&12&42.29 &  $-$64&50&33.76 &  3.10 &  2.3~$\pm$~0.5 &  1.0~$\pm$~0.5 &  0.01 &  $-$0.30 \\
   68 &  9&13&12.00 &  $-$64&49&12.30 &  3.28 &  3.6~$\pm$~0.8 &  0.8~$\pm$~0.6 &  $-$0.37 &  $-$0.02 \\
   69 &  9&10&59.37 &  $-$64&47&32.51 &  3.20 &  3.6~$\pm$~0.9 &  0.7~$\pm$~0.5 &  $-$0.17 &  $-$0.66 \\
   70 &  9&10&53.55 &  $-$65&01&35.59 &  3.12 &  6.2~$\pm$~1.4 &  2.7~$\pm$~1.4 &  0.09 &  0.12 \\
   71 &  9&10&43.16 &  $-$64&53&13.82 &  3.41 &  4.1~$\pm$~1.0 &  1.6~$\pm$~0.8 &  0.01 &  0.12 \\
   72 &  9&12&46.84 &  $-$65&04&05.62 &  3.41 &  2.7~$\pm$~0.7 &  1.1~$\pm$~0.9 &  $-$0.62 &  $-$0.51 \\
   73 &  9&13&25.04 &  $-$64&54&44.82 &  3.29 &  3.7~$\pm$~0.9 &  1.0~$\pm$~0.7 &  $-$0.21 &  0.45 \\
   74 &  9&10&40.75 &  $-$65&01&06.34 &  3.35 &  3.4~$\pm$~1.1 &  0.5~$\pm$~0.8 &  $-$0.48 &  $-$1.00 \\
   75 &  9&12&25.96 &  $-$64&42&14.74 &  3.32 &  4.4~$\pm$~1.1 &  1.2~$\pm$~0.8 &  $-$0.32 &  $-$0.34 \\
   76 &  9&12&13.28 &  $-$64&44&30.85 &  3.22 &  3.3~$\pm$~0.8 &  2.0~$\pm$~0.9 &  0.34 &  0.45 \\
   77 &  9&12&23.75 &  $-$64&57&10.31 &  3.17 &  2.5~$\pm$~0.7 &  0.9~$\pm$~0.7 &  $-$0.38 &  0.40 \\
   78 &  9&10&01.72 &  $-$64&49&35.37 &  3.61 &  10.1~$\pm$~2.2 &  4.3~$\pm$~2.1 &  $-$0.62 &  0.76 \\
   79 &  9&13&32.54 &  $-$65&01&54.65 &  3.23 &  4.1~$\pm$~1.0 &  3.7~$\pm$~1.9 &  $-$0.53 &  0.49 \\
   80 &  9&13&25.92 &  $-$64&41&54.45 &  3.14 &  6.4~$\pm$~1.6 &  2.3~$\pm$~1.6 &  $-$0.37 &  0.19 \\
   81 &  9&10&26.40 &  $-$64&58&32.21 &  3.20 &  4.5~$\pm$~1.2 &  1.4~$\pm$~1.1 &  0.33 &  $-$0.15 \\
   82 &  9&13&06.79 &  $-$64&46&17.01 &  3.30 &  3.6~$\pm$~0.9 &  1.3~$\pm$~0.6 &  0.18 &  $-$0.03 \\
   83 &  9&10&39.50 &  $-$64&51&40.97 &  3.17 &  4.1~$\pm$~1.0 &  1.9~$\pm$~1.0 &  $-$0.06 &  0.33 \\
   84 &  9&11&18.08 &  $-$64&44&05.44 &  3.30 &  2.7~$\pm$~0.8 &  0.5~$\pm$~0.6 &  $-$1.00 &  0.33 \\
   85 &  9&12&13.60 &  $-$64&47&53.24 &  3.07 &  2.1~$\pm$~0.5 &  1.2~$\pm$~0.6 &  0.40 &  0.28 \\
   86 &  9&12&29.41 &  $-$64&45&58.43 &  3.21 &  4.1~$\pm$~0.8 &  2.2~$\pm$~0.9 &  1.00 &  0.38 \\
   87 &  9&13&03.27 &  $-$64&43&57.14 &  3.30 &  5.1~$\pm$~1.2 &  2.6~$\pm$~1.2 &  0.96 &  0.24 \\
   88 &  9&11&21.58 &  $-$64&56&55.59 &  3.13 &  2.9~$\pm$~0.7 &  1.6~$\pm$~0.7 &  0.67 &  0.54 \\
   89 &  9&10&55.54 &  $-$65&02&46.31 &  3.32 &  6.2~$\pm$~1.5 &  2.4~$\pm$~1.4 &  0.21 &  0.07 \\
   90 &  9&14&02.85 &  $-$64&57&02.65 &  3.46 &  1.8~$\pm$~0.6 &  1.6~$\pm$~1.0 &  $-$0.45 &  $-$0.53 \\
   91 &  9&12&30.90 &  $-$64&54&28.17 &  3.35 &  1.1~$\pm$~0.3 &  1.2~$\pm$~0.6 &  0.97 &  $-$0.12 \\
   92 &  9&09&56.63 &  $-$64&48&46.89 &  3.11 &  4.7~$\pm$~1.5 &  0.4~$\pm$~1.0 &  $-$0.94 &  0.00 \\
\end{longtable}
}

\onllongtab{3}{
\begin{longtable}{cc@{$^h$}c@{$^m$}c@{$^s$~~~~}c@{\degr}c@{\arcmin}c@{\arcsec~~~~}c@{\arcsec~~~~}cccccc}
\caption{\label{table:4372src}
\object{NGC~4372} X-ray source properties. The columns contain the ID of the source, the position (RA$_{2000}$ and Dec$_{2000}$) with the error, the rate [$\times$~10$^{-3}$~count~s$^{-1}$], the flux converted with ECFs in band 0.5--10~keV [$\times$~10$^{-14}$~erg~cm$^{-2}$~s$^{-1}$] and the hardness ratios ($HR$). The $HR$ for two energy band $B1$ and $B2$ are defined by $HR=(B2-B1)/(B1+B2)$, $HR1$ for 0.5--1.5 and 1.5--3~keV energy bands, $HR2$ for 1.5--3 and 3--10~keV energy bands.} \\
\hline\hline
 ID & \multicolumn{3}{c}{RA$_{2000}$} & \multicolumn{3}{c}{Dec$_{2000}$} & \multicolumn{1}{l}{Error} & Rate & Flux & $HR1$ & $HR2$ \\
\hline
\endfirsthead
\caption{continued.}\\
\hline\hline
 ID & \multicolumn{3}{c}{RA$_{2000}$} & \multicolumn{3}{c}{Dec$_{2000}$} & \multicolumn{1}{l}{Error} & Rate & Flux & $HR1$ & $HR2$ \\
\hline
\endhead
\hline
\endfoot
    1 &  12&25&33.82 &  $-$72&27&47.25 &  2.74 &  221.0~$\pm$~8.4 &  74.8~$\pm$~9.5 &  $-$0.75 &  $-$0.72 \\
    2 &  12&25&11.18 &  $-$72&27&03.41 &  2.74 &  250.6~$\pm$~10.1 &  113.0~$\pm$~15.7 &  $-$0.62 &  $-$0.44 \\
    3 &  12&25&38.57 &  $-$72&49&23.94 &  2.81 &  38.1~$\pm$~3.3 &  10.5~$\pm$~4.7 &  $-$0.94 &  $-$1.00 \\
    4 &  12&26&24.22 &  $-$72&32&37.25 &  2.89 &  27.3~$\pm$~3.2 &  19.4~$\pm$~6.0 &  $-$0.05 &  $-$0.37 \\
    5 &  12&26&47.98 &  $-$72&42&15.92 &  3.02 &  15.8~$\pm$~2.3 &  9.1~$\pm$~4.1 &  $-$0.47 &  $-$0.28 \\
    6 &  12&28&01.38 &  $-$72&43&15.82 &  2.92 &  19.4~$\pm$~2.5 &  41.9~$\pm$~7.7 &  $-$0.01 &  0.17 \\
    7 &  12&28&09.69 &  $-$72&31&41.34 &  3.03 &  26.3~$\pm$~3.9 &  60.1~$\pm$~12.2 &  0.19 &  0.14 \\
    8 &  12&27&56.17 &  $-$72&41&04.30 &  3.03 &  13.4~$\pm$~2.2 &  23.6~$\pm$~6.2 &  $-$0.03 &  $-$0.07 \\
    9 &  12&25&27.76 &  $-$72&26&52.07 &  3.69 &  21.8~$\pm$~5.5 &  13.0~$\pm$~11.1 &  $-$0.10 &  $-$0.55 \\
   10 &  12&27&32.84 &  $-$72&40&13.53 &  3.32 &  5.1~$\pm$~1.3 &  2.9~$\pm$~2.6 &  $-$0.73 &  $-$1.00 \\
\end{longtable}
}

\end{document}